\documentclass[12pt]{iopart}
\usepackage{amsfonts}                   
\usepackage{amssymb}
\usepackage{amsbsy}
\usepackage{graphicx}
\usepackage{cite}

\begin{document}
\review[Pair-contact process with diffusion]
{The non-equilibrium phase transition of the pair-contact process with 
diffusion}
\author{Malte Henkel$^a$ \ and Haye Hinrichsen$^b$} \vglue 3mm
\address{$^a$\ Laboratoire de Physique des Mat{\'e}riaux,\footnote{Laboratoire 
associ\'e au CNRS UMR 7556}
         Universit{\'e} Henri Poincar{\'e} Nancy I, B.P. 239, 
	 F--54506 Vand{\oe}uvre l{\`e}s Nancy Cedex, France} \vglue 1mm
\address{$^b$\ Fakult\"at f\"ur Physik und Astronomie, 
         Universit\"at W\"urzburg, D--97074 W\"urzburg, Germany}


\begin{abstract}
The pair-contact process $2A\to 3A$, $2A\to\emptyset$ with diffusion of 
individual particles is a simple branching-annihilation processes which
exhibits a phase transition from an active into an absorbing phase 
with an unusual type of critical behaviour which had not been seen before. 
Although the model has attracted considerable interest during the past few 
years it is not yet clear how its critical behaviour can be characterized 
and to what extent the diffusive pair-contact process represents an independent 
universality class. Recent research is reviewed 
and some standing open questions are outlined. 
\end{abstract}


\noindent Date: the 16$^{\rm th}$ of February 2004/
                the 7$^{\rm th}$ of May 2004\\
\submitto{\JPA} 
\maketitle
\parskip 2mm


\newcommand{\BEQ}{\begin{equation}}     
\newcommand{\BEA}{\begin{eqnarray}}
\newcommand{\EEQ}{\end{equation}}       
\newcommand{\EEA}{\end{eqnarray}}
\def\be{\begin{equation}}
\def\ee{\end{equation}}
\def\ba{\begin{eqnarray}}
\def\ea{\end{eqnarray}}
\newcommand{\eps}{\varepsilon}          
\newcommand{\vph}{\varphi}              
\newcommand{\vth}{\vartheta}            
\newcommand{\D}{{\rm d}}                
\newcommand{\II}{{\rm i}}               
\renewcommand{\Re}{{\rm Re\ }}          
\renewcommand{\Im}{{\rm Im\ }}          
\newcommand{\arcosh}{{\rm arcosh\,}}    
\newcommand{\erf}{{\rm erf\,}}          
\newcommand{\wit}[1]{\widetilde{#1}}    
\newcommand{\lap}[1]{\overline{#1}}     
\newcommand{\rar}{\rightarrow}          
\newcommand{\ket}[1]{{\left| #1 \right>}} 
\newcommand{\bra}[1]{{\left< #1 \right|}} 
\newcommand{\braket}[2]{{\left< #1 \left| #2 \right.\right>}}
\renewcommand{\vec}[1]{\boldsymbol{#1}} 
\newcommand{\zeile}[1]{\vskip #1 \baselineskip} 
\newcommand{\vekz}[2]
     {\mbox{${\begin{array}{c} #1  \\ #2 \end{array}}$}}
\newcommand{\appsection}[2]{\setcounter{equation}{0} \section*{Appendix #1. #2}
\renewcommand{\theequation}{#1\arabic{equation}}
              \renewcommand{\thesection}{#1} }

\newcommand{\é}{\'e}                	
\newcommand{\è}{\`e}
\newcommand{\à}{\`a}
\newcommand{\ù}{\`u}


\def\xvec{{\vec{x}}}		        
\def\nupar{{\nu_\parallel}}		
\def\nuperp{{\nu_\perp}}		
\def\text{\mbox}			


{\small \input{revue_PCPD.inh} }

\section{Introduction and History} \label{sect1}

The {\it `pair-contact process with diffusion'} (PCPD) is a classical 
reaction-diffusion process describing stochastically 
interacting particles of a single species which react spontaneously 
whenever two of them come into contact. In its simplest version the 
PCPD involves two competing reactions, namely
\begin{eqnarray*}
\mbox{fission:}      \qquad && 2A \to 3A\,, \\
\mbox{annihilation:} \qquad && 2A \to \emptyset\,.
\end{eqnarray*}
In addition individual particles are allowed to diffuse. 
Moreover, depending on the
specific variant under consideration, an exclusion principle 
or a similar mechanism 
may impose a restriction such that the particle density cannot diverge.

During the past four years the PCPD has attracted a lot of attention. 
The interest in this model is first motivated by the fact that the PCPD 
exhibits a non-equilibrium phase transition caused by the
competing character of fission and annihilation. In the so-called
{\it active phase}, the fission process  dominates and the system 
evolves towards a fluctuating steady-state 
characterized by a certain stationary particle density $\rho_s>0$. 
Contrarily, in the {\it absorbing phase} the annihilation process dominates 
and the density of particles decreases continuously until the system reaches
a so-called {\em absorbing state} (for example the empty lattice) from where it 
cannot escape. 
Since absorbing states can only be reached but not be left, 
the process is generically out of equilibrium. 
The active and the absorbing phase are separated by a continuous 
phase transition with a non-trivial critical behaviour. 

As we shall see, despite the simplicity of the PCPD, the 
properties of its phase transition(s) are quite involved and rather
surprisingly, at the time of writing no clear consensus on its critical
behaviour has been achieved. This makes the PCPD an ideal laboratory to 
explore the most advanced techniques from either field-theory or simulation. 
On the other hand, due to the intensive recent research, the PCPD 
offers the possibility to introduce all these techniques as applied 
to the same model and thereby to critically appreciate their respective
strengths and weaknesses. 

The classification of non-equilibrium phase transitions from an active 
phase into an inactive phase with absorbing states is a challenging problem 
of contemporary statistical physics~\cite{Marr99,Hinr00}. 
Generally it is believed that such transitions belong to a 
finite number of universality classes with steady-state properties being
characterized by a quartet $(\beta,\beta',\nu_\perp,\nu_\parallel)$ of
critical exponents and certain universal scaling functions.\footnote{Only
if a certain duality symmetry holds, the {\it a priori} different \cite{Mend94}
steady-state exponent $\beta$ of the order parameter and 
$\beta'$ of the survival probability will agree.} So far only few universality 
classes are firmly established, the most important ones being directed 
percolation (DP)~\cite{Harr74,Kinz83,Ligg85,Odor04}, 
the parity-conserving (PC) class 
of branching-annihilating random walks~\cite{Zhon95,Card96}, $Z_2$-symmetric 
transitions in the voter model~\cite{Dorn01,Droz03}, the general epidemic 
process~\cite{Card85,Jans85}, and absorbing phase transitions coupled to 
a conserved field~\cite{Ross00,Lueb02,Lueb03}. The current interest in the 
PCPD is motivated by the perspective 
that it may represent yet another independent 
universality class of non-equilibrium phase transitions which had not 
been studied before. The asymptotic critical behaviour, however, seems to 
be masked by strong corrections to scaling so that it is not yet entirely 
clear how the transition in the PCPD should be characterized and classified. 
The purpose of this review is to summarize the present state of knowledge 
and to point out open questions. 

\begin{figure}
\centerline{\includegraphics[width=130mm]{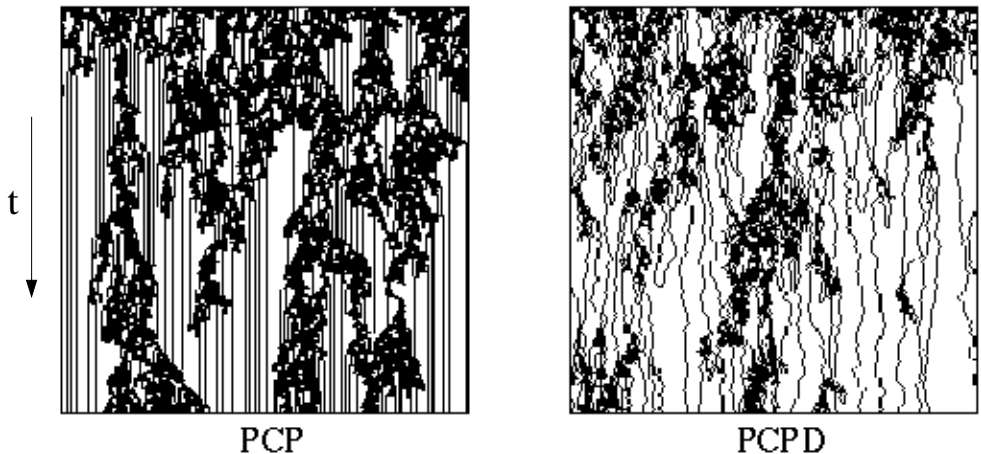}}
\caption{\label{FIGPCP}
One-dimensional pair-contact process starting with a fully occupied lattice 
at criticality. Left: In the pair-contact process without diffusion (PCP) 
solitary particles cannot diffuse, leading to frozen patterns of separated 
vertical lines. Right: In the PCPD, where individual particles are allowed 
to diffuse, offspring production can be restarted after long times when two 
diffusing particles meet, leading to a very different visual appearance of 
the process.
}
\end{figure}

The first investigation of this class of models considered the 
\textit{non-diffusive} pair-contact process (PCP)~\cite{Jens93,Jens93b}
with at most one particle per site. 
In this case individual particles are completely immobile 
so that activity can only spread in space by offspring 
production at empty nearest-neighbour sites. 
On a one-dimensional lattice, the PCP may be implemented by fission 
$\emptyset AA/AA\emptyset\to AAA$ and annihilation 
$AA\to\emptyset\emptyset$. A typical space-time plot of a critical PCP 
is shown in the left panel of figure~\ref{FIGPCP}. Since solitary 
particles do not diffuse, any configuration of spatially separated 
particles is frozen. Therefore, on an infinite lattice the model 
has infinitely many absorbing states. More specifically, on a 
one-dimensional lattice with $L$ sites and either periodic or 
free boundary conditions the model has $2^L$ states of which
\BEQ \label{pcp:nst}
N_{\rm stat}(L) \simeq N_{0}\, g_+^L \;\; , \;\; 
N_0 = \left\{ \begin{array}{ll} 1          & \mbox{\rm ~~;~ periodic b.c.} \\
                                1.17\ldots & \mbox{\rm ~~;~ free b.c.}
\end{array} \right.
\EEQ
are absorbing \cite{Carl01}, where $g_+=(1+\sqrt{5})/2\simeq 1.618$ is the 
golden mean (see appendix~A). Numerical simulations in both
one dimension ($1D$) \cite{Jens93,Dick98} and in $2D$ \cite{Kamp99} have shown 
that the steady-state transition in the non-diffusive PCP belongs to DP.
Although a rigorous proof is not yet available, recent numerical high-precision
studies of universal scaling functions in various dimensions
support this claim~\cite{Lueb02b}. The result is remarkable since the 
PCP does not obey the assumptions of the so-called 
DP-conjecture~\cite{Gras79,Jans81}, which states that under certain generic 
conditions any non-equilibrium phase transition into a {\em single} 
absorbing state should belong to DP. 
Spreading from a single seed, however, may show non-universal properties
depending on the density and correlations of the surrounding
particles~\cite{Mend94,Lope97,Jime03}.
To explain the observed critical behaviour in the
PCP several field-theoretic approaches have been suggested~\cite{Muno96,Wijl02}.
%
%
%
%

In the PCPD, where in addition to the binary PCP reactions individual particles 
are also allowed to diffuse, the situation is fundamentally different. 
Unlike the PCP without particle diffusion, the PCPD has only two absorbing 
states, $N_{\rm stat}=2$. These are the empty lattice and a homogeneous state 
with a single diffusing particle. 
Therefore, and in contrast to spreading processes 
such as DP, where diffusion 
does not alter the critical behaviour~\cite{Dick89}, introducing diffusion 
in the pair-contact process should be regarded as a {\em singular perturbation} 
which changes the structure of the absorbing phase. In fact, even the visual 
appearance changes drastically, as shown in the right panel of 
figure~\ref{FIGPCP}. 

The pair-contact process {\em with} diffusion was already suggested in 1982 by
Grassberger~\cite{Gras82}, who expected a critical 
behaviour ``distinctly different'' 
from DP.\footnote{He rather considered the process 
$2A\to 3A$, $3A\to A$, which is believed to exhibit the 
same type of phase transition as 
the PCPD.} 15 years later the problem was rediscovered, under the name of 
``annihilation-fission process'', by 
Howard and T{\"a}uber~\cite{Howa97}, who proposed a bosonic 
field-theory for the one-dimensional PCPD. Because of the bosonic nature of
their model, the particle density is unrestricted and thus diverges in the 
active phase. Furthermore,  since their field-theory turned out to be 
non-renormalizable, no quantitative statements on 
the phase transition between the active and the inactive phase could be made. 
This triggered one of us (M.H.), together with J.F.F. Mendes, to attempt a
non-perturbative study of the $1D$ PCPD, formulated on a discrete lattice and 
with at most one particle per site. Starting from the master equation rewritten
as a matrix problem of an associated quantum Hamiltonian $H$, the longest
relaxation time was obtained from the real part of the lowest eigenvalue 
of $H$ and steady-state observables were found by forming matrix elements with 
the right ground state of $H$. The results were compared with the ones of
a Monte Carlo simulation. While the estimates of the location of the transition
were in reasonable agreement between the two methods, it turned out that for 
the longest chains ($L=21$ sites) accessible, finite-size estimates for the 
critical exponents could not yet be reliably extrapolated towards the
$L\to\infty$ limit~\cite{MendesUnpublished}. As we shall see, even today 
the difficulties encountered in that first study 
are not yet completely overcome. 

At that time, the likely solution to these difficulties 
appeared to be the investigation of larger lattices 
by adapting White's density-matrix renormalization group (DMRG) 
\cite{Whit92,Whit93} to non-equilibrium systems with non-hermitian matrices $H$
(for a recent review, see \cite{Scho04}).
Kaulke and Peschel \cite{Kaul98} used the DMRG to study biased diffusion, which
leads to a $q$-symmetric Hamiltonian $H$ which is similar to a symmetric
matrix. The feasibility of the Hamiltonian approach using the DMRG for truly 
non-equilibrium systems without detailed balance was demonstrated with 
E. Carlon and U. Schollw\"ock, on a model in the DP class \cite{Carl99}. 
Equipped with this technique, we then returned to the investigation of a 
lattice model of the PCPD with an exclusion principle by density-matrix 
renormalization group methods \cite{Carl01}. This paper was followed by a long 
series of numerical and analytical studies 
\cite{Hinr01a,Odor00,Hinr01b,Odor01,Park01,Henk01a,Henk01b,Odor02a,Odor02b,Noh04,Park02b,Dick02b,Hinr02a,Kock02,Odor02c,Paes03b,Bark03,Hinr03a} 
and released the still ongoing debate concerning the asymptotic critical 
behaviour of the PCPD at the transition. Until now a surprising variety of 
possible scenarios has been proposed, the main viewpoints being that the 
active-absorbing phase transition in the PCPD
\begin{enumerate}
\item[-] should represent a novel universality class with a unique set of 
critical exponents~\cite{Hinr01a,Henk01b,Park02b,Kock02},
\item[-] may represent {\em two different} universality classes 
depending on the diffusion rate~\cite{Odor00,Odor02c} and/or the number of
space dimensions \cite{Paes03b},
\item[-] can be interpreted as a cyclically coupled DP and annihilation 
process~\cite{Hinr01b}, 
\item[-] could be regarded as a marginally perturbed DP process with 
continuously varying critical exponents~\cite{Noh04},
\item[-] may have exponents depending continuously on the diffusion constant
$D$ \cite{Dick02b,OdorPrivat},
\item[-] may cross over to DP after very long time~\cite{Hinr02a,Bark03}, and
\item[-] is perhaps related to the problem of non-equilibrium wetting in 1+1 
dimensions~\cite{Hinr03a}.
\end{enumerate}
Given these widely different and partially contradicting 
conclusions, it is clear that considerably more
work will be needed to understand the behaviour of this so simple-looking
model. In this review, we shall give an introduction to the PCPD, review
the research done on it and discuss these suggested scenarios in detail. 
We thereby hope to stimulate further research in order to finally understand
the properties of this system. 

\section{Phenomenological scaling properties}\label{sect2}

It is well-known that systems in the directed percolation universality 
class are usually characterized by the so-called 
DP-conjecture~\cite{Jans81,Gras82} which specifies certain 
conditions for DP critical behaviour. Similarly,  
experience suggests that the critical phenomenon observed in the PCPD
is not restricted to a particular model, rather it is expected to appear in a
large variety of models which are thought to be characterized by the 
following phenomenological features:
\begin{enumerate}
\item Particles of a single species diffuse in a $d$-dimensional space.
\item The phase transition is driven by two competing {\em binary} 
reactions for 
particle creation and removal, i.e., two particles have to come into contact 
in order to be eliminated or to produce offspring.
\item There is a finite number of absorbing states of which at least one 
is reachable.
\item There are no unusual features such as quenched randomness or 
unconventional symmetries.
\end{enumerate} 
This means that the characteristic critical behaviour of the PCPD
is also expected to exist in many other models, including e.g. 
generic reaction-diffusion processes of the form
\begin{eqnarray*}
(a) \quad && 2A \to 3A, \ \ 2A \to \emptyset \\
(b) \quad && 2A \to 3A, \ \ 2A \to A \\
(c) \quad && 2A \to 4A, \ \ 2A \to A \\
(d) \quad && 2A \to 3A, \ \ 3A \to \emptyset 
\end{eqnarray*}
All these models differ significantly from both DP and PC processes
in two respects: Firstly, processes in the DP class are unary processes, 
i.e. isolated particles 
can decay or trigger branching. The paradigmatic example is the contact process
\cite{Harr74,Ligg85} with the reactions $A\to2A$, $A\to\emptyset$,
exclusion on the lattice, and possibly
diffusion of single particles. In contrast, in the PCPD single diffusing 
particles cannot react unless they collide in pairs. 
Secondly, in a typical system in
the PC class such as $2A\to\emptyset$, $A\to 3A$, the parity of the total 
number of particles is conserved 
and there is at least one unary reaction while this is not so in the PCPD. 
In particular, all PCPD-like models have an inactive (or absorbing) phase, 
where either the annihilation process $2A\to \emptyset$ or else the coagulation 
process $2A\to A$ dominates. Therefore, in the inactive phase of the PCPD 
and related models 
the density $\rho(t)$ of particles decays algebraically as~\cite{Tous83}
\begin{equation}
\rho(t) \sim
\left\{
\begin{array}{ll}
t^{-d/2}     & \text{ for } d<2 \\
t^{-1} \ln t & \text{ for } d=2 \\
t^{-1}       & \text{ for } d>2
\end{array}
\right. .
\end{equation}
This behaviour is quite distinct from the time-dependence of the 
particle density for systems in the DP universality class, where the decay 
of $\rho(t)$ is exponential in time.

Studying the PCPD one has to distinguish between 
two fundamentally different types 
of models, namely, the {\em unrestricted} PCPD, where the number 
of particles per site is not constrained, and the {\em restricted}  
PCPD, where the particle number on any given site 
is effectively bounded.\footnote{These two variants
are often referred to as {\em bosonic} and {\em fermionic} models. 
This nomenclature,
however, is misleading since e.g. models with a soft constraint are restricted
but bosonic.} This restriction can be implemented 
either as a hard constraint by an exclusion principle or as a soft constraint 
in form of a higher-order process for particle removal (such as 
$3A\to \emptyset$), which prevents the particle density from diverging. 
The difference between the two types is most pronounced in the active phase,
where in the unrestricted case the particle density diverges 
while it saturates at a finite value in restricted variants. 
Since phase transitions with a well-defined particle density
in the active phase are perceived as more natural, 
most studies consider the restricted PCPD.

In all recent studies of the restricted PCPD it is assumed that the 
critical behaviour at the transition can be described in terms of the 
same scaling laws as found in other absorbing phase transitions, 
although possibly with a different set of critical exponents and 
scaling functions. For example, in a Monte Carlo simulation one usually
measures the temporal evolution of the particle density 
starting from a homogeneous 
initial state (usually a fully  occupied lattice). 
In this case the particle density 
$\rho$ is expected to obey the asymptotic scaling form
\begin{equation}
\rho(t,L,\epsilon) = t^{-\delta} \ 
f_\rho\left(t^{1/z}/L,\epsilon t^{1/\nupar}\right) \ ,
\end{equation} 
where $L$ is the lateral system size, $\epsilon=p-p_c$ measures the distance 
from criticality, and $f_{\rho}$ is a scaling function with an 
appropriate asymptotic behaviour.
Here and in what follows a scaling limit
$t\to\infty$, $L\to\infty$ and $\epsilon\to 0$ is implied. 
The exponent $\delta=\beta/\nupar$ describes the 
power-law decay of the density at criticality while $z=\nupar/\nuperp$ 
is the dynamical critical exponent. Apart from metric factors
the scaling function and the critical exponents are determined by 
the universality class of the phase transition and the boundary conditions.

Alternatively one can start the process with an active seed, 
i.e. in the present case with a {\it pair} of particles at the origin. 
As usual in this type of simulation~\cite{Gras79}, 
one studies the survival probability $P_s$ which scales as
\begin{equation}
P_s(t,L,\epsilon) = t^{-\delta'} \ 
f_{P_s}\left(L t^{1/z},\epsilon t^{1/\nupar}\right) \ ,
\end{equation} 
where $\delta'=\beta'/\nupar$. Similarly, the average number of 
particles averaged over all runs scales as
\begin{equation}
N(t,L,\epsilon) = t^{\eta} \ 
f_N\left(L t^{1/z},\epsilon t^{1/\nupar}\right) \ ,
\end{equation} 
and the mean-square spreading from the origin averaged over surviving runs is
\begin{equation}
R^2(t,L,\epsilon) = t^{2/z} \ 
f_{R^2}\left(L t^{1/z},\epsilon t^{1/\nupar}\right) \ .
\end{equation} 
Here $\delta'=\beta'/\nupar$ is an independent exponent, although in the DP 
and the PC class the two exponents $\beta$ and $\beta'$ are equal.\footnote
{In DP the equality is caused by a time-reversal symmetry \cite{Gras79} while 
for the PC class it is so far only observed numerically.}
The so-called initial-slip exponent $\eta$ is related 
to the other exponents by the generalized 
hyperscaling relation~\cite{Mend94,Muno97}
\begin{equation} \label{HYPER}
\eta = \frac{d}{z}-\delta-\delta' \ .
\end{equation}
The definition of the survival probability in the 
PCPD is a little subtle. Clearly, the presence of a single particle as a 
criterion for survival is not sufficient since then $P_s(t)$ would 
tend to a constant. A natural definition, however, appears to be to consider a
run as surviving as long the system has not yet reached one of the two 
absorbing states, i.e. there have to be at least two particles in 
the system. A stronger condition is to require the presence of at least
one {\em pair} in the system as advocated in \cite{Dick02b,Noh04}, 
see sections 4.3 and 5.2. 

According to standard scaling theory the quartet of exponents 
$(\beta,\beta',\nuperp,\nupar)$ characterizes the universality class 
of the transition. For DP and PC transitions, where $\beta=\beta'$, 
the remaining three independent exponents in one spatial dimension 
have the values
\begin{equation}
\!\!\!\!\!\!\!\!\!\{\beta,\nuperp,\nupar\} \approx
\left\{ \begin{array}{ll}
\{0.276486(8), \ 1.096854(4), \  1.733847(6)\}       
&\text{ for DP~\cite{Jens99}} \\
\{0.922(5), \hspace{6mm} \ 1.85(3), \hspace{8mm} \ 3.22(6)\}  
&\text{ for PC~\cite{Zhon95}}
\end{array}
\right. 
\end{equation} 
The central open question about the PCPD is whether it belongs 
to one of the known classes or whether it represents a novel universality 
class with an independent set of critical exponents. 
   
\section{Mean-field approaches}\label{sect3}

Mean-field theories are a convenient tool in order to obtain a first
orientation about the possible critical behaviour of a statistical system. 
To be specific, we shall formulate the PCPD as a lattice model, where each
site may either be empty or be occupied by a single particle. The model
evolves by random-sequential dynamics according to the rules~\cite{Carl01}
\begin{eqnarray}
AA\emptyset,\,\emptyset AA \rightarrow AAA  && \qquad \text{with rate }  
(1-p)(1-D)/2 
\nonumber \\
AA \rightarrow \emptyset\emptyset           && \qquad \text{with rate } p(1-D) 
\label{pcpd:raten} \\
A\emptyset \leftrightarrow\emptyset A       && \qquad \text{with rate }\,  D 
\nonumber \ ,
\end{eqnarray}
where $0\leq p \leq 1$ and $0\leq D\leq 1$ are the control parameters. We are
therefore treating a restricted version of the PCPD and shall comment on
the unrestricted version as discussed in~\cite{Howa97} as necessary. 

All mean-field treatments must at some stage neglect some of the correlations
present in a model. A systematic way of constructing a sequence of improving 
mean-field theories was devised by ben-Avraham and K\"ohler \cite{BenA92} and
extending earlier work \cite{Guto87,Dick88}. 
In their so-called $(n,m)$-approximation they work with clusters of $n$ 
sites. The approximation is made by factorizing the probabilities for 
configurations which occupy more than $n$ 
sites in terms of $n$-site probabilities but in a way that there is an
overlap of $m$ sites between adjacent clusters. 
More specifically, if $\mathfrak{A,B,C}$ are such overlapping 
clusters, and if $P(\mathfrak{A}|\mathfrak{B})=P(\mathfrak{A})/P(\mathfrak{B})$
is the conditional probability to find $\mathfrak{A}$ 
for given $\mathfrak{B}$, the approximation 
$P(\mathfrak{ABC})\approx P(\mathfrak{AB})P(\mathfrak{BC}|\mathfrak{B})$ is
used. In this way, correlations up to $n$ sites are taken into account. 
It is assumed throughout that the cluster probabilities $P(\mathfrak{A})$ 
are spatially translation-invariant. 

For example, consider a five-site cluster in the state $ABCDE$. In the
$(3,1)$- and $(3,2)$-approximations, respectively, the probability 
$P(ABCDE)$ of this cluster is expressed as
\BEA
\hspace{-2.0truecm}P^{(3,1)}(ABCDE) &\approx& 
P(ABC)P(CDE|C) = \frac{P(ABC)\,P(CDE)}{P(C)} 
\nonumber \\
\hspace{-2.0truecm}P^{(3,2)}(ABCDE) &\approx& 
P(ABC)P(BCDE|BC) 
\approx P(ABC)\frac{P(BCD)P(CDE|CD)}{P(BC)}\hspace{1truecm} 
\\
&=& \frac{P(ABC)\,P(BCD)\,P(CDE)}{P(BC)\,P(CD)} \,.
\nonumber
\EEA
Therefore, in the context of the $(3,m)$-approximation, 
the basic variables have to
be constructed from the independent three-sites probabilities. 

It has been argued that the $(n,n-1)$-approximations are qualitatively the
most reliable ones~\cite{BenA92} and this conclusion has been generally 
accepted. In particular, the simplest of these schemes 
is the $(1,0)$-approximation, which is also 
referred to as the {\em site approximation} ($N=1$) 
or as {\em simple mean-field}. 
Similarly, the $(2,1)$-approximation is called the {\em pair approximation} 
($N=2$) and the $(3,2)$- and $(4,3)$-approximations are referred to as 
{\em triplet} ($N=3$) and {\em quartet} approximations ($N=4$), respectively.

In what follows we shall denote an occupied site by $\bullet$ and an
empty site by~$\circ$. A three-site cluster $AA\emptyset$ will then be
denoted as $\bullet\bullet\circ$ and has the probability 
$P(AA\emptyset)=P_{\bullet\bullet\circ}$. For notational simplicity, we 
shall only consider the one-dimensional case in this section. 

\subsection{Simple mean-field approximation for the particle density}

The simplest mean-field approach considers only the single-site probabilities. 
If we let $\rho(t)=P_{\bullet}(t)$ and also assume that it is space-independent,
the rate equation for $\rho(t)$ is easily derived. In table~\ref{app:site}
we collect those reactions from (\ref{pcpd:raten}) 
which change the mean particle-density, together
with the change $\Delta N_{\bullet}$ in the number of occupied sites 
$N_{\bullet}$ by this reaction. 

\begin{table}
\footnotesize
\begin{center}
\begin{tabular}{|c|ll|} \hline
reaction & $\Delta N_{\bullet}$ & rate  \\ \hline
$\bullet\bullet\circ\to\bullet\bullet\bullet$ 
         & +1                   & $\frac{1}{2}(1-p)(1-D)\rho^2(1-\rho)$ \\
$\circ\bullet\bullet\to\bullet\bullet\bullet$ 
         & +1                   & $\frac{1}{2}(1-p)(1-D)\rho^2(1-\rho)$ \\
$\bullet\bullet \to \circ\circ$ 
         & -- $2$                 & $p(1-D)\rho^2$ \\ \hline         
\end{tabular} 
\caption{\label{app:site} Rates in the site approximation of the PCPD.}
\end{center}
\end{table} 
%
Adding the contributions to $\dot{\rho}$ and rescaling time $t\to t(1-D)^{-1}$ 
one easily finds
\BEQ \label{N1approx}
\frac{\D\rho(t)}{\D t} = 
(1-p)\, \rho(t)^2\Bigl( 1 - \rho(t)\Bigr) - 2p\, \rho(t)^2
\EEQ
which implies the following large-time behaviour of the 
mean density \cite{Carl01}
\BEQ \label{N1app:tgross}
\rho(t) \simeq \left\{ \begin{array}{ll} 
\frac{1-3p}{1-p}+\mathfrak{a}\exp(-t/\tau) & \mbox{\rm ~~;~ if $p<1/3$}\\[2mm]
\sqrt{3/4}\cdot t^{-1/2} & \mbox{\rm ~~;~ if $p=1/3$}\\[2mm]
(3p-1)^{-1}\cdot t^{-1} & \mbox{\rm ~~;~ if $p>1/3$}
\end{array} \right.
\EEQ
where $\tau=(1-p)/(1-3p)^2$ and $\mathfrak{a}$ is a constant which depends on
the initial conditions. We see that the site approximation reproduces the
intuitive expectation of a phase transition between an active phase, where
$\rho(t)\to\rho_{\infty}>0$ for large times, and an absorbing phase, where
$\rho(t)$ vanishes in the $t\to\infty$ limit. We also see that throughout the
absorbing phase, the approach towards the steady-state of vanishing particle
density is algebraic, although mean-field theory is not capable of reproducing
the correct decay $\rho(t)\sim t^{-1/2}$ which one would obtain in one 
dimension if the fluctuation effects neglected here were taken into account. 

{}From the point of view of a continuum theory it is natural to introduce a 
coarse-grained particle density $\rho(\xvec,t)$ and to postulate a suitable 
rate equation. A reasonable rate equation should be given by\footnote{We warn 
the reader that simple rate equations may not at all be adequate for the case 
of an unrestricted density, see \cite{Houc02} and section 6.} 
\begin{equation} \label{champmoyen}
\frac{\partial}{\partial t} \rho(\xvec,t) = b \rho^2(\xvec,t) - 
c \rho^3(\xvec,t) + D \nabla^2 \rho(\xvec,t) \,,
\end{equation}
and we expect for a restricted model with the rates
(\ref{pcpd:raten}) that $b=1-3p$ and $c=1-p$, see eq.~(\ref{N1approx}). 
On the other hand, for unrestricted models without any constraint on the 
occupation of a space point $\vec{x}$, one would have $c=0$~\cite{Howa97}
so that the mean particle density $\rho(\vec{x},t)$ diverges in the
active phase $b>0$. Hence the cubic term with $c>0$ can be 
interpreted to mean that a newly created particle requires
some empty space to be put into. In later sections, we shall discuss
variants of the PCPD with a soft constraint, generating effectively 
such a cubic term with $c>0$. 
As we shall see this will lead to important consequences. 

At first sight, one might believe that eq.~(\ref{N1approx}) contains with
respect to (\ref{champmoyen}) the further approximation that the 
space-dependence of $\rho(\vec{x},t)\mapsto \rho(t)$ is neglected. Still, 
both eqs.~(\ref{N1approx}) and (\ref{champmoyen}) lead to the same long-time
behaviour of the density. To see this, choose some spatial domain
$\Omega$ and consider the spatially averaged density
\BEQ
\overline{\rho}(t)=
\frac{1}{|\Omega|}\int_{\Omega}\!\D\vec{x}\,\rho(\vec{x},t)\,,
\EEQ
where $|\Omega|$ denotes the volume of $\Omega$ and $\rho(\vec{x},t)$ solves
(\ref{champmoyen}). Now, if $c>0$ and
$\overline{\rho}(0)>0$ and with the boundary condition 
$\nabla\rho|_{\partial\Omega}=0$, it can be shown
rigorously \cite{Henk03} that there exists a constant $b'$ so that $b'/b>0$, 
and a positive constant $c'$, which are both independent of $|\Omega|$ 
and of $\overline{\rho}(0)$ such that the inequalities 
\BEA
~~~b \overline{\rho}(t)^2 - c'\overline{\rho}(t)^3 \leq
&\frac{\D\overline{\rho}(t)}{\D t}& \leq 
~~~b' \overline{\rho}(t)^2 - c\overline{\rho}(t)^3 \;\; ; \;\;
\mbox{\rm ~~if $b\geq 0$} 
\nonumber \\
-|b'| \overline{\rho}(t)^2 - c'\overline{\rho}(t)^3 \leq
&\frac{\D\overline{\rho}(t)}{\D t}& \leq 
-|b| \overline{\rho}(t)^2 - c\overline{\rho}(t)^3 \;\; ; \;\;
\mbox{\rm ~~if $b\leq 0$} 
\EEA
hold true. In addition, one has in the quadratic mean the 
convergence $\rho(\vec{x},t)\to\overline{\rho}(t)$ for large times 
\cite{Henk03}. Therefore, the time-dependence of the averaged density 
$\overline{\rho}(t)$ can be bounded both from above and from below by a 
solution of the form (\ref{N1app:tgross}). Hence we can conclude that the 
space-dependence of $\rho(\vec{x},t)$ is not essential for the long-time 
behaviour of the mean density, rather the crucial approximation in mean-field 
theories is made in neglecting correlations between the states of different 
sites. 

Finally, a standard dimensional analysis of eq. (\ref{champmoyen}), 
see \cite{Hinr00}, yields the mean-field critical exponents
\begin{equation}
\label{MFExponents}
\beta^{MF}=1,\quad 
\nuperp^{MF}=1,\quad
\nupar^{MF}=2.
\end{equation}
As will be discussed in section~\ref{FTSec}, 
these exponents are expected to 
be valid in $d \geq  2$ dimensions. In fact, high-precision simulations in 2D 
performed by {\'O}dor, Marques and Santos \cite{Odor02a} confirm this 
prediction within numerical errors and up to logarithmic factors 
for various values of the diffusion rate. 
In one spatial dimension, however, fluctuation effects are expected to be 
relevant, leading to different values. A better description of these
fluctuation effects requires to consider larger clusters, going
beyond the site approximation. These approaches will be 
discussed in the following subsections.
   
\subsection{The pair approximation: Two transitions ?}

\begin{table}
\footnotesize
\begin{center}
\begin{tabular}{|c|lll|l|} \hline
reaction & $\Delta N_{\bullet}$ & $\Delta N_{\bullet\bullet}$ & rate & \\ \hline
$\bullet\bullet\circ\circ\to\bullet\circ\bullet\bullet$ 
         & ~~0 & -- $1$ & $D uvw/\rho(1-\rho)$ & $\times 2$ \\
$\circ\bullet\circ\bullet\to\circ\circ\bullet\bullet$ 
         & ~~0 & +1     & $D v^3/\rho(1-\rho)$ & $\times 2$ \\ \hline
$\bullet\bullet\bullet\bullet\to\bullet\circ\circ\bullet$ 
         & -- $2$ & -- $3$ & $p(1-D)u^3/\rho^2$ & \\
$\bullet\bullet\bullet\circ\to\bullet\circ\circ\circ$ 
         & -- $2$ & -- $2$ & $p(1-D)u^2v/\rho^2$ & $\times 2$ \\
$\circ\bullet\bullet\circ\to\circ\circ\circ\circ$ 
         & -- $2$ & -- $1$ & $p(1-D)uv^2/\rho^2$ & \\ \hline
$\bullet\bullet\circ\bullet\to\bullet\bullet\bullet\bullet$ 
         & +1 & +2 & $\frac{1}{2}(1-p)(1-D) uv^2/\rho(1-\rho)$ & $\times 2$ \\
$\bullet\bullet\circ\circ\to\bullet\bullet\bullet\circ$ 
         & +1 & +1 & $\frac{1}{2}(1-p)(1-D) uvw/\rho(1-\rho)$ & $\times 2$ \\
\hline
\end{tabular} 
\caption{\label{app:pair} Rates in the pair approximation of the $1D$ PCPD. 
In the last column, a symmetry factor coming from parity-symmetry is included.}
\end{center}
\end{table} 
%
Treating the $1D$ PCPD in the pair approximation, we assume again that the
probabilities $P(AB)$ are translation-independent and furthermore 
that the system is left/right-invariant, 
viz. $P_{\bullet\circ}(t)=P_{\circ\bullet}(t)$. Then,
because of $P_{\bullet}=P_{\bullet\circ}+P_{\bullet\bullet}$ and
$P_{\bullet\bullet}+2P_{\bullet\circ}+P_{\circ\circ}=1$, there are two 
independent variables which may be chosen as the particle density 
$\rho(t)=P_{\bullet}(t)$ and the pair density $u(t)=P_{\bullet\bullet}(t)$. 

We now illustrate the standard methods~\cite{BenA92,Marr99} how to find the 
equations of motions for $\rho(t)$ and 
$u(t)$ in the pair approximation. Introduce the shorthand
$v=P_{\bullet\circ}=\rho-u$ and $w=P_{\circ\circ}=1-2\rho+u$. The reactions
changing the number $N_{\bullet}$ of occupied sites of the number 
$N_{\bullet\bullet}$ of occupied pairs are listed with their rates 
in table~\ref{app:pair}, grouped into the separate contributions of the
three reactions.
In the last group, we only need to take into account those reactions $2A\to 3A$ 
which modify the particle configuaration on a given site. Adding the 
contributions to $\dot{\rho}$ and $\dot{u}$, their equations of motion are 
easily derived~\cite{Carl01}:
\BEA
\dot{\rho}(t) &=& -2(1-D)p\,u(t)+(1-D)(1-p)\,
\left(\rho(t)-u(t)\right)\frac{u(t)}{\rho(t)} 
\\
\dot{u}(t) &=& -(1-D)p\,u(t)\frac{2u(t)+\rho(t)}{\rho(t)} -2D
\frac{(\rho(t)-u(t))(u(t)-\rho(t)^2)}{\rho(t)(1-\rho(t))} 
\nonumber \\
& &+(1-D)(1-p)\frac{(\rho(t)-u(t))(1-u(t))u(t)}{\rho(t)(1-\rho(t))}
\EEA
The pair approximation is the lowest-order cluster approximation which allows
the effects of diffusion to be treated explicitly. 
In particular, in the $D\to 1$
limit, the site approximation eq.~(\ref{N1approx}) is recovered for 
$\rho(t)$, while the pair density $u(t)=\rho(t)^2$. 

The critical line is again at $p_c(D)=\frac{1}{3}$ for $D\geq\frac{1}{7}$, but
if $0\leq D<\frac{1}{7}$, one has $p_c(D)=\frac{1}{5}(1+3D)/(1-D)$. In addition,
the scaling of both $\rho(t)$ and of $u(t)$ -- and similarly of their
steady-state solutions $\rho_{\infty}$ and $u_{\infty}$ -- does depend on 
whether $D>\frac{1}{7}$ where $u\sim\rho^2$ or whether $D<\frac{1}{7}$ where
$u\sim \rho$. The steady-state particle density $\rho_{\infty}\sim p_c(D)-p$ 
for all $D$, and if $D\ne\frac{1}{7}$ one has $u(t)\sim t^{-1}$ along the
critical line. These results might suggest the existence
of two distinct universality classes along the critical line, but since the
pair approximation is the simplest cluster mean-field theory which
permits to study the effect of particle diffusion at all, this conjecture
certainly needs further verification. 

%
%
\begin{figure}
\centerline{\includegraphics[width=110mm]{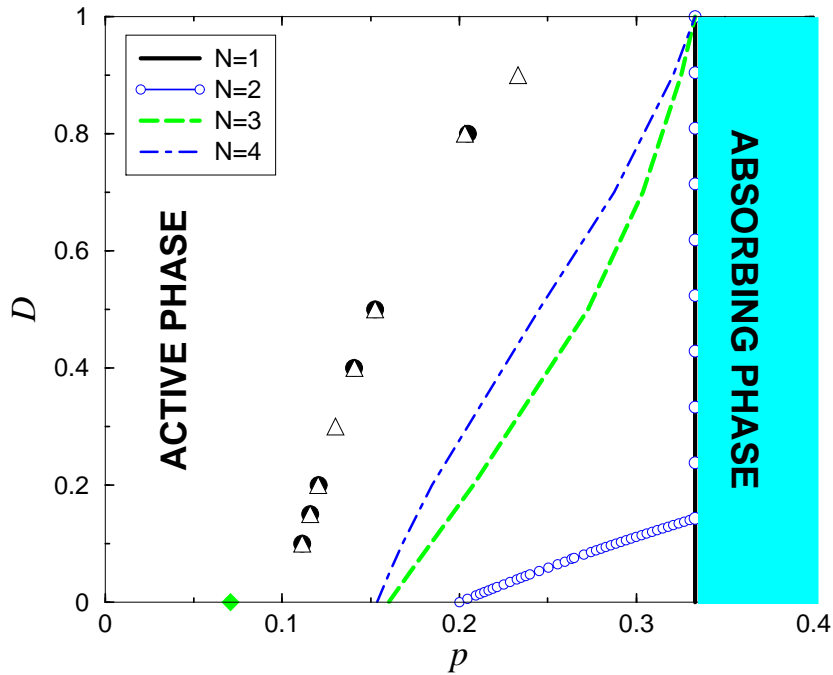}}
\caption[Phase diagram]{
Phase diagram of the one-dimensional restricted PCPD with the rates as defined 
in {\protect eq.~(\ref{pcpd:raten})}, after {\protect\cite{Odor02c}}. 
The figure shows the boundary between the active 
and the absorbing phases according to cluster mean-field theory, ranging
from site ($N=1$) to quartet approximation ($N=4$). 
The filled dots give values of $p_c(D)$ as found from the 
DMRG {\protect\cite{Carl01}} and the open triangles those found in
Monte Carlo simulations {\protect\cite{GrasPrivat}}. The grey diamond gives
the critical point $p_c(0)$ of the PCP {\protect\cite{Kamp99}}.
\label{PhasePCPD}}
\end{figure}
%
%
In figure~\ref{PhasePCPD} the phase diagram of the $1D$ PCPD is shown. 
The pair approximation ($N=2$) certainly is an improvement as 
compared to simple mean-field ($N=1$) but is still quite far from the exact 
location of $p_c(D)$ as obtained from DMRG calculations or Monte-Carlo 
simulations which will be described in section~\ref{sect4}.

\subsection{Higher cluster approximations}

In the pair approximation of the PCPD, the particle and
pair densities in the steady-state 
are related by $u_{\infty}=\rho_{\infty}(1-3p)/(1-p)$. This
means that the particle and pair densities have to vanish simultaneously which
is certainly not true for the simple PCP without diffusion ($D=0$), see
section~\ref{sect1}. This failure of the pair approximation can be overcome 
within the triplet approximation ($N=3$). Consider first the PCP where  
$D=0$ \cite{Marq01,Lueb02b}. Again assuming left/right symmetry, 
the independent variables
may be chosen as $P_{\bullet\bullet\bullet}$, $P_{\bullet\bullet\circ}$,
$P_{\bullet\circ\circ}$ and $P_{\bullet\circ\bullet}$. Now the critical
point occurs at $p_c(0)\simeq 0.128$, where the steady-state pair density 
$u_{\infty}\sim p_c(0)-p$ vanishes linearly, but the critical particle density 
$\rho_{\infty,c}\simeq 0.23$ remains finite. These results
have been extended up to the sextet approximation ($N=6$) \cite{Dick02a} 
and further to $N=12$ \cite{Szol02} and remain 
qualitatively valid.   

It is then natural to extend this study to the whole phase diagram. This was
done in \cite{Odor02c}, including the quartet approximation ($N=4$).
As can be seen in figure~\ref{PhasePCPD} the kink in $p_c(D)$ observed in 
the pair approximation is absent for larger clusters. 
In addition, along the critical line, the
steady-state particle density $\rho_{\infty}\sim p_c(D)-p$ for all values of
$D>0$, whereas $u_{\infty}\sim (p_c(D)-p)^2$ and this holds for both
$N=3$ and $N=4$. This result strongly suggests that 
there is only one universality 
class along the critical line \cite{Odor02c}. 
This finding is in agreement with 
simulational results, see section~\ref{sect4}. 
A similar result had already been obtained before in the 2D process 
$2A\to\emptyset$, $2A\to 4A$ with single-particle diffusion \cite{Odor02a}. 

To summarize, the apparent presence of two universality classes for $D>0$
should be an artifact of the pair approximation -- at least in one
dimension -- and it has already
been observed earlier that in some cases, cluster mean-field theories
may indeed yield qualitatively incorrect phase diagrams \cite{BenA92}. 
Generically, in the PCPD some
reactions involve three neighbouring sites, and furthermore the explicit
consideration of at least three neighbouring sites is required in order to
become sensitive to the large number of absorbing states. Therefore it appears 
natural that cluster mean-field approximations which 
already neglect some of the
correlations present in the reaction terms themselves are unlikely to be 
reliable (in fact, the site approximation may be too simplistic 
to yield anything but a single transition). 
Rather, the pair approximation is good enough to reveal the presence of a
PCP transition for $D=0$ which is distinct from the one which occurs for $D>0$,
but since three-sites correlations are not treated correctly, the PCP
transition might have been stabilized artificially even for some values $D>0$. 
Similar considerations should apply to any interacting particle system where 
the interactions occur through the contact of at least two particles
(see also section 6). 

\section{Numerical studies of the critical behaviour in one dimension} 
\label{sect4}

\subsection{Density-matrix renormalization group study}  

Since the first non-perturbative study of the PCPD \cite{Carl01} was made using
the reformulation of the problem through the quantum hamiltonian formalism, we
shall begin with a brief description of that method, 
see \cite{Schu00,Hinr00,Henk03a}
for recent reviews. The starting point is provided by the master equation, 
cast into the form
\BEQ \label{mastereq}
\partial_t \ket{P(t)} = - H \ket{P(t)} \,,
\EEQ
where $\ket{P(t)} = \sum_{\{\sigma\}} P(\{\sigma\};t)\ket{\{\sigma\}}$ is the
state vector of the system at time $t$, $\ket{\{\sigma\}}$ is the state
vector of a particle configuration $\{\sigma\}$ on the lattice, and
$P(\{\sigma\};t)$ is the time-dependent probability of that configuration. 
For a chain of $L$ sites with a maximal occupancy of one particle per site
the {\em quantum Hamiltonian} $H$ is a $2^L\times 2^L$ matrix with elements
\BEQ
\bra{\{\sigma\}}H\ket{\{\tau\}} = - w(\tau\to\sigma) + \delta_{\sigma,\tau}
\sum_{\{\upsilon\}} w(\tau\to\upsilon)\,,
\EEQ
where $w(\tau\to\sigma)$ is the transition rates from the configuration
$\{\tau\}$ to the configuration $\{\sigma\}$. 
Since the elements of the columns of $H$ add up to zero, $H$ describes
indeed a stochastic process. From the condition of probability conservation
$\sum_{\{\sigma\}} P(\{\sigma\};t) =1$ it follows that
$\bra{\mathfrak{s}}H = 0$,
where $\bra{\mathfrak{s}}=\sum_{\{\sigma\}}\bra{\{\sigma\}}$ is a left 
eigenvector of $H$ with eigenvalue 0. Since the particle reactions are 
generically irreversible, $H$ is in general non-hermitian and therefore it has 
distinct right and left eigenvectors, which we denote by 
$\ket{0_r},\ket{1_r},\ldots,\ket{n_r}$ and 
$\bra{0_l}, \bra{1_l},\ldots,\bra{n_l}$, respectively with the corresponding
`energy levels' $E_0, E_1,\ldots, E_n$ ordered according to
$0 = E_0 \leq \Re E_1\leq \cdots \Re E_n$. Clearly, $\braket{n_l}{m_r}=0$
if $E_n\ne E_m$ and we normalize states such that $\braket{n_l}{n_r}=1$. 
We also set $\bra{\mathfrak{s}}=\bra{0_l}$. 

Given the rates (\ref{pcpd:raten}) of the PCPD, the associated quantum
Hamiltonian $H$ on a chain with $L$ sites is easily constructed. The 
stationary states of the model correspond to right eigenvectors of $H$ with
vanishing eigenvalue. For the PCPD with $D>0$ these are, for
both periodic and free boundary conditions~\cite{Carl01}, the two
absorbing states
\BEA
\ket{0_r} &:=& \ket{\circ\circ\circ\cdots\circ}  \nonumber \\
\ket{1_r} &:=& \frac{1}{L}\left(\ket{\bullet\circ\circ\cdots\circ} 
+ \ket{\circ\bullet\circ\cdots\circ} + \cdots 
+ \ket{\circ\circ\circ\cdots\bullet}\right) \,.
\EEA
Consequently, the first two eigenvalues $E_0$ and $E_1$ 
vanish so that the inverse 
relaxation time towards the  absorbing state is 
given by the smallest `energy gap'
$\Gamma =\Gamma(p,D;L) := \Re E_2$. Finally, formally solving the master 
equation (\ref{mastereq}), the average value 
$\langle\mathfrak{O}\rangle(t)$ of
an observable $\mathfrak{O}$ is given by
\BEQ
\langle\mathfrak{O}\rangle(t) = \bra{\mathfrak{s}}\mathfrak{O}\ket{P(t)}
=\sum_n \bra{\mathfrak{s}}\mathfrak{O}\ket{n_r}\braket{n_l}{P(0)} e^{-E_n t}\,,
\EEQ
where we tacitly assumed the completeness (and biorthogonality) of the right 
and left eigenvector systems. Although the quantum hamiltonian formalism was
originally introduced with the aim of relating non-equilibrium systems on
infinitely long chains to integrable quantum systems, we are interested here
in the numerical analysis of systems on {\em finite} chains with $L$ sites.
Then the long-time behaviour is given by the smallest inverse relaxation
time $\tau=\xi_{\|}=\Gamma^{-1}$ with the following finite-size behaviour
\BEQ \label{GammaL}
\Gamma \sim \left\{\begin{array}{ll}
\exp(-L/\xi_{\perp}) & \mbox{\rm ~~;~ if $p<p_c(D)$} \\
L^{-z}               & \mbox{\rm ~~;~ if $p=p_c(D)$} \\
L^{-2}               & \mbox{\rm ~~;~ if $p>p_c(D)$} 
\end{array} \right.
\EEQ
where $z=\nu_{\|}/\nu_{\perp}$ and $\xi_{\perp}\sim|p-p_c(D)|^{-\nu_{\perp}}$
is the spatial correlation length. For large lattices and $p\geq p_c(D)$, 
one expects that the lowest energies should satisfy a massless dispersion 
relation of the form $E(k)\sim k^z$, and since the lowest momenta should scale 
as $k_{\rm min}\sim L^{-1}$, the phenomenological scaling (\ref{GammaL}) is
recovered. 

Having found $\Gamma(p,D;L)$ on a sequence of finite lattices, one proceeds
as usual, see \cite{Priv90} for a collection of reviews. The 
critical parameters (here $p_c(D)$ and $z$) are extracted by forming
the logarithmic derivative
\BEQ \label{logderGamma}
Y_L(p,D) := \frac{\ln[\Gamma(p,D;L+1)/\Gamma(p,D;L-1)]}{\ln[(L+1)/(L-1)]} \,.
\EEQ
However, since the entire absorbing phase is critical and therefore
$\lim_{L\to\infty}\Gamma(p,D;L)=0$ 
the habitual method of looking for the intersection
of two curves $Y_L(p,D)$ and $Y_{L'}(p,D)$ cannot be applied \cite{Carl01}. 
Rather, the critical point $p_c(D)$ must be found by fixing
$d$ and $L$ and finding the value of $p=p_L(D)$ which maximizes $Y_L(p,D)$
(see Appendix~B for further details). Then the sequence of estimates has to
be extrapolated to $p_c(D)=\lim_{L\to\infty} p_L(D)$. Finally,
for $L$ large, one expects
\BEQ \label{4:gl:Exponenten}
Y_L(p,d) \simeq \left\{\begin{array}{ll}
-L/\xi_{\perp} & \mbox{\rm ~~;~ if $p<p_c(D)$} \\
-z             & \mbox{\rm ~~;~ if $p=p_c(D)$} \\
-2             & \mbox{\rm ~~;~ if $p>p_c(D)$} 
\end{array} \right.
\EEQ
and one finally obtains $z=-\lim_{L\to\infty} Y_L(p_c(D),D)$ \cite{Carl01}. 
In addition an estimate of the order parameter exponent 
$\beta/\nu_{\perp}$ can be obtained from the
steady-state density profile at $p=p_c(D)$ 
\BEQ
\rho_{L}(\ell) = \bra{\mathfrak{s}} \hat{n}(\ell)\ket{0_r} 
= L^{-\beta/\nu_{\perp}} f(\ell/L)\,,
\EEQ
where $\hat{n}(\ell)$ is a particle number operator at site $\ell$ and $f$
is a scaling function. The double degeneracy of the ground state was lifted
by adding a particle creation process $\emptyset\to A$ with a small rate $p'$. 
For estimates of the bulk exponent, one may set $\ell=L/2$ and 
$\rho_L(L/2)\sim L^{-\beta/\nu_{\perp}}$ as expected. On the other hand, 
if one sets $\ell=1$, in the limit of $p'\to 0$ one has the scaling form 
$\rho_{L}(1)\sim L^{(\beta_1-\beta)/\nu_{\perp}}$ from which the surface 
critical exponent $\beta_1/\nu_{\perp}$ can be 
determined \cite{Carl99}.\footnote{For models with a non-degenerate ground
state the use of a surface rate may be avoided
by estimating the profile from $N(\ell)=\bra{1_l}\hat{n}(\ell)\ket{1_r}$.
At criticality, the finite-size scaling forms 
$N(L/2)\sim L^{-\beta/\nu_{\perp}}$ and
$N(1)\sim L^{-\beta_1/\nu_{\perp}}$ are expected \cite{Laur98,FHL98,Carl99}.} 

The lowest eigenvalue $\Gamma$ and the corresponding eigenvector are
found from the non-symmetric DMRG algorithm \cite{Carl99,Carl01} on an
open chain. In comparison with Monte-Carlo simulations, the DMRG method 
as described here directly yields the steady-state with a high numerical 
precision, does not suffer from a critical slowing-down in the vicinity of 
the critical point and for non-disordered systems does not require the use 
of a random-number generator. On the other hand, the number of lattice sites 
which can be treated is relatively restricted. 
For example, in \cite{Carl01} data for $\Gamma_L$ with $L\lesssim 30$ sites and 
for $\rho_L$ with $L\leq 48$ sites could be obtained. Larger lattices with
up to $L=60$ sites have been obtained more recently \cite{Henk01b,Bark03}. 
In the absorbing phase, however, the finite-size sequences converge very well 
and the accuracy of the estimated limits can be considerably enhanced through 
the {\sc bst} extrapolation algorithm \cite{Buli64,Henk88}. The expected
dynamical and order-parameter exponents $z=2$ and $\beta/\nu_{\perp}=1$ are 
recovered to high accuracy. On the other hand, along the critical
line, the required finite-size extrapolations are affected by some 
ill-understood correction terms which preclude the use of sequence 
extrapolation algorithms. The final results for $p_c(D)$ and the 
exponents $z$ and $\beta/\nu_{\perp}$ are collected in table~\ref{ExpTable}.
Since results for several values of $D$ were obtained, a discussion of
systematic effects on the estimates becomes possible. 

While the values of $\beta/\nu_{\perp}$ are almost independent of $D$, the
results for $z$ show a considerable variation which was argued in \cite{Carl01}
to come from a subtle finite-size correction. Taking mean values and comparing
with the values of the other universality classes available at the time,
namely directed percolation (DP) and the parity-conserved (PC) universality
classes, the values of $z$ and $\beta/\nu_{\perp}$ of \cite{Carl01} are far 
from those of DP but intriguingly close to those of PC, 
see table~\ref{ExpTable}. Because of this surprising 
coincidence the possibility
that the active-inactive transition of the PCPD might belong to the PC 
universality class emerged, in spite of the fact that there is no obvious 
counterpart for the parity conservation law in the PCPD, i.e. there is no 
symmetry separating the dynamics into two sectors~\cite{Hinr01a}. 
In fact, in both cases the inactive phase is characterized by an algebraic 
decay of the density as $t^{-1/2}$. Moreover, both models have two absorbing 
states. Still, the unambiguous identification of a steady-state universality
class requires the determination of four independent exponents, see
\cite{Mend94,Marr99,Hinr00}, rather than the two obtained in \cite{Carl01}. 

Indeed, small-scale Monte-Carlo simulations \cite{Hinr01a,Odor00} seemed to 
confirm the DMGR results for $p_c(D)$ and, for $D$ not too large, also the 
exponents $z$ and $\beta/\nuperp$. However,
either the exponent bound $\beta<0.67$ \cite{Hinr01a} or the estimate
$\beta\simeq 0.58$ \cite{Odor00} was significantly smaller than the 
expected PC value $0.92$. In addition, the effective exponents $\delta'$ and 
$\eta$ in seed simulations were shown to differ significantly from the expected 
PC values. These results rather suggested that
the PCPD might be a novel nonequilibrium universality class,
distinct from that of PC as well as DP processes.

\subsection{High-precision Monte-Carlo simulations}   

\begin{table}
\footnotesize
\begin{center}  
\begin{tabular}{|l||l|l|l|l|l|l|l|l|}
\hline
&&&&&&&&\\
Reference & \multicolumn{1}{c|}{$D$} & \multicolumn{1}{c|}{$p_c$} & 
\multicolumn{1}{c|}{$\delta$} & 
\multicolumn{1}{c|}{$z=\frac{\nu_{\|}}{\nu_{\perp}}$} & 
\multicolumn{1}{c|}{$\beta$} & 
\multicolumn{1}{c|}{$\beta/\nuperp$} & \multicolumn{1}{c|}{$\delta'$} & 
\multicolumn{1}{c|}{$\eta$} \\
&&&&&&&&\\
\hline \hline
Carlon \etal \cite{Carl01}
& 0.1	& 0.111(2) & --		& 1.87(3)	& --	& 0.50(3) & --& --\\
& 0.15	& 0.116(2) & --		& 1.84(3)	& --	& 0.49(3) & --& --\\
& 0.2	& 0.121(3) & --		& 1.83(3)	& --	& 0.49(3) & --& --\\
& 0.35	& 0.138(1) & --		& 1.72(3)	& --	& 0.47(3) & --& --\\
& 0.5	& 0.154(1) & --		& 1.70(3)	& --	& 0.48(3) & --& --\\
& 0.8	& 0.205(1) & --		& 1.60(5)	& --	& 0.51(3) & --& --\\
\hline
Hinrichsen \cite{Hinr01a}
& 0.1	& 0.1112(1) & 0.25(2)	& 1.83(5)	& $<0.67$ & 0.50(3) & 0.13(2) & 0.13(2)\\
\hline
\'Odor \cite{Odor00} 	
& 0.05 & 0.25078 & 0.273(2) 	& -- 		& 0.57(2) 	& --& 0.004(6)& 0.10(2)\\     
& 0.1 	 & 0.24889 & 0.275(4) 	& -- 		& 0.58(1) 	& --& --      & --     \\
& 0.2 	 & 0.24802 & 0.268(2) 	& -- 		& 0.58(1) 	& --& 0.004(6)& 0.14(1)\\
& 0.5 	 & 0.27955 & 0.21(1) 	& -- 		& 0.40(2) 	& --& 0.008(9)& 0.23(2)\\
& 0.9 	 & 0.4324  & 0.20(1) 	& -- 		& 0.39(2) 	& --& 0.01(1) & 0.48(1)\\
\hline
\'Odor \cite{Odor01} 	
& 0.2 	 & 0.17975(8)& 0.263(9) & -- 		& 0.57(1) 	& -- & --& --\\
& 0.4 	 & 0.2647(1) & 0.268(8) & -- 		& 0.58(1) 	& -- & --& --\\
& 0.7 	 & 0.3528(2) & 0.275(8) & -- 		& 0.57(1) 	& -- & --& --\\
\hline
Park \etal \cite{Park01}
& * & 0.0895(2) & 0.236(10) 	& 1.80(2)	& 0.50(5) & -- &  $\approx 0.1$ & $\approx 0.2$\\     
\hline
Noh \& Park \cite{Noh04}
& 0.1 & 0.1112(1) & 0.27(4) 	& 1.8(2)	& 0.65(12) & 0.50(5) & 0.09(2) & 0.18(5) \\     
\hline
Park \& Kim \cite{Park02b}
A & * & 0.03081(4) & 0.241(5) 	& 1.80(10)	& 0.519(24) & -- &  0.11(3) & 0.15(3)\\     
Park \& Kim \cite{Park02b}
B & * & 0.28735(5) & 0.242(5) 	& 1.78(5)	& 0.496(22) & -- &  0.13(3) & 0.15(3)\\     
\hline
Dickman 
& 0.1	 & 0.10648(3) & 0.249(5) & 2.04(4) & 0.546(6)	& 0.503(6) & --& --\\
\& Menezes \cite{Dick02b} 
& 0.5	 & 0.12045(3) & 0.236(3) & 1.86(2) & 0.468(2)	& 0.430(2) & --& --\\
& 0.85 & 0.13003(1) & 0.234(5) & 1.77(2) & 0.454(2)	& 0.412(2) & --& --\\
\hline
Hinrichsen \cite{Hinr02a}
& 1 & 0.29998(4) & 0.22(1) & 1.78(5) & -- & -- & 0.13(1) & 0.22(3) \\     
\hline
Kockelkoren    
& *	& * & 0.200(5) 	& 1.70(5) 	& 0.37(2) 	& -- & -- & --\\
\& Chat\'e \cite{Kock02} & &  & &       &               &    &    &   \\
\hline
\'Odor \cite{Odor02c,OdorPrivat} 	
&0.05 & 0.10439(1) & 0.216(9) & 2.0(2)  & 0.411(10) & 0.53(7) & -- & -- \\
&0.1  & 0.10688(1) & 0.206(7) & 1.95(1) & 0.407(7)  & 0.49(2) & -- & -- \\
&0.2  & 0.11218(1) & 0.217(8) & 1.95(1) & 0.402(8)  & 0.46(3) & -- & -- \\
&0.5  & 0.13353(1) & 0.206(7) & 1.84(1) & 0.414(16) & 0.41(2) & -- & -- \\
&0.7  & 0.15745(1) & 0.214(5) & 1.75(1) & 0.39(1)   & 0.38(2) & -- & -- \\
\hline
Barkema  
& 0.1 & 0.11105(5) & 0.17     & --      & -- & --      & -- & -- \\
\& Carlon \cite{Bark03} 
& 0.2 & --         & 0.17     & 1.70(3) & -- & 0.28(4) & -- & -- \\	
& 0.5 & 0.15245(5) & 0.17(1)  & --      & -- & 0.27(4) & -- & -- \\
& 0.9 & 0.2335(5)  & 0.17     & 1.61(3) & -- & --      & -- & -- \\
\hline
\multicolumn{9}{|c|}{}\\
\multicolumn{9}{|c|}{\bf Related models:}\\
\multicolumn{9}{|c|}{}\\
\hline
Hinrichsen \cite{Hinr01b}
& 1 & 0.6929(1) & 0.21(1) & 1.75(10) 	& 0.38(6) 	& --& 0.15(1)& 0.21(2)\\     
\hline
\'Odor \cite{Odor02b} 	
& * & 0.3253(1) & 0.19(1) & 1.81(2) 	& 0.37(2) 	& -- & -- & -- \\     
\hline
Hinrichsen \cite{Hinr02b}
& * & * & 0.21(2) & 1.75(2) 	& -- 	& --& -- & --\\     
\hline
\multicolumn{9}{|c|}{}\\
\multicolumn{9}{|c|}{\bf Known universality classes:}\\
\multicolumn{9}{|c|}{}\\
\hline
DP \cite{Jens99}
& &  & 0.1595 & 1.5807 & 0.2765 & 0.2521 &	0.1595 & 0.3137 \\
\hline
PC \cite{Zhon95} 
& & & 0.286(2) & 1.74(1)	& 0.922(5)	& 0.497(5) & 0 & 0.286(2)\\	
\hline
\end{tabular}
\vglue 5mm
\caption{\label{ExpTable}
Summary of estimates for the critical exponents of the $1D$ PCPD. 
A star indicates that the control parameter is effectively fixed or defined in 
a different way. Several variants 
of models with differently defined critical thresholds $p_c(D)$ were used.}
\end{center}  
\end{table}

The first simulational study of the PCPD with sequential dynamics 
(the rates (\ref{pcpd:raten}) were chosen for maximal efficiency) yielded the 
critical-point estimates \cite{MendesPrivat}
\BEQ \label{pcD}
p_c(0.1) = 0.1105(5) \;\; , \;\;
p_c(0.5) = 0.153(1)
\EEQ
which are in good agreement with later results shown in figure~\ref{PhasePCPD} 
and table~\ref{ExpTable}. We point out that several inequivalent definitions
of the critical threshold $p_c(D)$ are in use and only the definitions of
the articles \cite{Carl01,Hinr01a,Noh04,Bark03} and of figure~\ref{AmplitudenV} 
-- which gives the results of \cite{GrasPrivat} -- are consistent with each 
other and eq.~(\ref{pcD}). A systematic high-precision Monte-Carlo simulation 
was performed by \'Odor \cite{Odor00}. In order to implement the model on a 
parallel computer, he investigated a variant of the PCPD with {\em synchronous 
updates} on three sublattices. Depending on the value of the diffusion rate he 
found continuously varying exponents $\delta$, $\beta$ and $\eta$ 
(see table~\ref{ExpTable}), concluding that the process does not belong to 
the PC universality class. Instead the value of $\beta$ seemed to jump 
from $\beta \approx 0.58$ for $0.05\leq D \leq 0.2$ to smaller values at 
$D>0.5$. This observation together with the aforementioned pair mean-field 
approximation led \'Odor to the conjecture that the PCPD might 
belong to \textit{two different} universality classes for low and high 
diffusion rate. Although it seems to be surprising how the choice of a 
rate can switch between two universality classes we note that a similar 
scenario has been observed in a certain two-species particle 
model~\cite{Wijl98}, although the specific mechanism in these models is
not obvious in the PCPD. We shall return to this problem in section~6. 

As pointed out in \cite{Hinr01a} the unusual critical behaviour of 
the PCPD should be seen in a large variety of binary reaction-diffusion 
processes with absorbing states, in particular in the coagulation-fission 
process $2A\to 3A$, $2A\to A$. Simulating the coagulation-fission 
process with branching sidewards
\begin{eqnarray}
AA\emptyset,\,\emptyset AA \rightarrow AAA  && \qquad \text{with rate }  
(1-p)(1-D)/2 \nonumber \\
AA \rightarrow \emptyset A/A\emptyset && \qquad  \text{with rate } p(1-D)/2 
\\
A\emptyset \leftrightarrow\emptyset A && \qquad \text{with rate }\,  D 
\nonumber \ 
\end{eqnarray}
as well as a symmetric variant with offspring production in the middle 
of two particles
\begin{equation}
A\emptyset A \rightarrow AAA  \qquad \qquad \ \ \text{with rate }  
\lambda (1-p)(1-D) \nonumber
\end{equation}
\'Odor \cite{Odor01} could confirm this expectation, provided that the 
diffusion rate is high enough. The observed exponents seemed to be in agreement 
with those of the PCPD for small diffusion rates, leading \'Odor to 
the conjecture that the coagulation-fission process may belong to a 
single universality class. 

The special case, where the rates for coagulation and diffusion are equal, 
i.e. $D=p(1-D)/2$, can be solved exactly for any value of $\lambda$ 
\cite{Henk01a}, using the 
well-known method of interparticle distribution functions~\cite{BenA90}. 
As pointed out in~\cite{Henk01a}, in this case the dynamics is special in 
so far as the fission process can only fill voids of an invariant skeleton 
of coagulating random-walks. Unfortunately, in this case the active phase 
is inaccessible so that the exact solution does not 
provide any information about 
the critical exponents of a possible universality class of PCPD transitions, 
rather the system always crosses over to a decay of the form 
$\rho(t)\sim t^{-1/2}$. Nevertheless the solution reveals that the 
finite-size scaling of the relaxation times does not depend on the 
fission rate and thereby allows to test a recent extension to 
non-equilibrium phase transitions \cite{Henk01b} of 
the Privman-Fisher finite-size scaling forms \cite{Priv84}.

Concerning the possibility of two universality classes or even continuously 
varying exponents, Park and Kim \cite{Park02b} showed by numerical simulations 
that in models, where the rates of diffusion and annihilation are tuned in such 
a way that the process without particle production would become exactly 
solvable, one obtains a well-defined set of critical exponents~\cite{Park02b} 
(see table~\ref{ExpTable}). In this sense, they argue, the transition in 
the PCPD may be considered as universal. Contrarily Dickman and 
Menezes~\cite{Dick02b} claim that the transition in 
the PCPD is non-universal. Their argument will be reviewed in the next 
subsection. 
%
%
\begin{figure}
\centerline{\includegraphics[width=155mm]{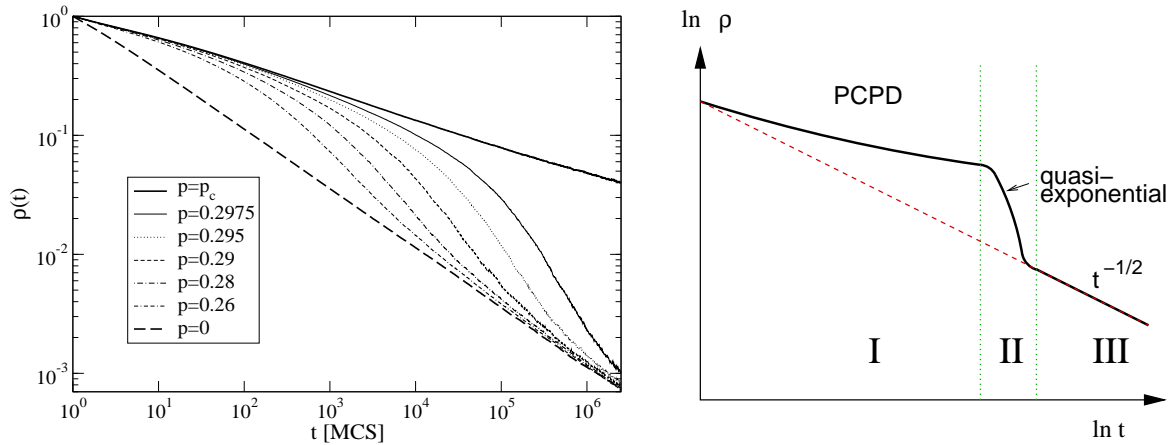}}
\caption[Inactive]{Left panel: Decay of the particle density in the 
inactive phase for different values of $p$ in comparision with the critical 
PCPD (bold line) and pure coagulation (dotted line).  
Right panel: Sketch of the sudden decay of the particle density in the
inactive phase (see text). After {\protect\cite{Hinr02a}}. 
\label{FigInactivePhase}}
\end{figure}
%
%

As a third possibility one of us (H.H.) suggested
an extremely slow crossover to DP~\cite{Hinr02a}. Simulating a 
cellular automation of the PCPD which was originally
introduced by Grassberger~\cite{GrasPrivat}, it was argued that the 
scaling regime is not reached, even after six decades in time. Instead 
the effective exponents display a slow drift as time proceeds. Remarkably, 
all exponents seem to drift in the direction of DP values so that one cannot 
rule out that the critical behaviour of the 1+1-dimensional PCPD may eventually 
tend to DP. According to these arguments the DP process manifests itself in the 
dynamics of pairs of particles, while the diffusing background of solitary 
particles becomes irrelevant as $t \to \infty$ (obviously, this mechanism 
should work exclusively in one spatial dimension while in higher dimensions the 
diffusive background may be increasingly relevant leading to an effective 
mean-field behaviour). Recently Barkema and Carlon \cite{Bark03} 
presented extrapolation results, arriving at a similar conclusion 
(see section~\ref{Extrapolation}).
Contrarily, a recent simulation of the PCPD with non-restricted occupancy per 
site performed by Kockelkoren and Chat\'e (see section~\ref{Kockelkoren}) 
does not show such a drift, supporting the hypothesis
of a single universality class distinct from previously known classes.

The DP hypothesis poses a conceptual problem, namely, how does the algebraic
decay $\rho(t) \sim t^{-1/2}$ in the inactive phase comply with DP, where the 
decay is exponential? In this context it is interesting to monitor the 
crossover of the particle density in a slightly subcritical PCPD. As shown in  
figure~\ref{FigInactivePhase}, there are three different 
temporal regimes. In the 
first regime (region I) the particle density decays very slowly as in a 
critical PCPD. After a characteristic time, however, the density 
decays rapidly (region II) until it crosses over to an algebraic decay 
$t^{-1/2}$ (region III). Remarkably, the amplitude of this asymptotic power 
law does {\em not} depend on $p-p_c$. Therefore, the breakdown in region II 
becomes more and more pronounced as the critical threshold is approached and 
might tend to a quasi-exponential decay in the limit $p \to p_c$.

\subsection{Universal moments and amplitudes}

Since the numerical values of the available exponents seem to vary 
between different universality classes, alternative diagnostic
tools might be of value. Following~\cite{Dick98}, 
let us consider the order-parameter moments  
\BEQ
m_n := \langle \rho^n\rangle
\EEQ
{}From the scaling arguments reviewed in section~\ref{sect2} it is known
that ratios such as $m_4/m_2^2$, $m_3/m_1^3$, $m_3/(m_1 m_2)$ or $m_2/m_1^2$ 
should be universal. In fact, moment ratios are just special values of
the scaling functions and hence their numerical values may be used, like
the critical exponents, to identify the universality class of a given 
model. Further one may consider the cumulants \cite{Dick98} 
\BEQ
K_2 := m_2 - m_1^2 \;\; , \;\;
K_4 := m_4 -4m_3m_1 -3m_2^2 +12m_2m_1^2-6m_1^4
\EEQ
whose ratios $K_4/K_2^2$ and $K_2/m_1^2=m_2/m_1^2-1$ should be 
universal as well. 

\begin{table}
\footnotesize
\begin{center}
\begin{tabular}{|c|c||lllll|c|} \hline
$d$ & model 
& $m_4/m_2^2$ & $m_3/m_1^3$ & $m_3/(m_1 m_2)$ & $m_2/m_1^2$ & $K_4/K_2^2$ 
& Ref. \\ \hline
1 & DP  & 1.554(2) & 1.526(3) & 1.301(3) & 1.1736(2) & -0.505(3) & \cite{Dick98} \\
  & PCP & 1.558(2) & 1.529(3) & 1.303(3) & 1.1738(2) & -0.493(3) & \cite{Dick98} \\ 
  & PC  &          &          &          & 1.3340(4) &           & \cite{Dick02b} \\ \hline
2 & DP  & 2.093(8) & 2.080(1) & 1.569(1) & 1.3257(5) & -0.088(4) & \cite{Dick98} \\
  & PCP & 2.07(1)  & 2.067(9) & 1.56(1)  & 1.323(3)  &           & \cite{Kamp99} \\ \hline
\end{tabular} 
\caption{\label{Momentenverhaeltnisse}
Some universal moment and cumulant ratios at criticality for the  
directed percolation universality class (DP), the pair-contact process (PCP) 
and the parity-conserving class (PC) for an infinite lattice in $d=1,2$ 
dimensions.}
\end{center}
\end{table} 

For future reference, we reproduce in table~\ref{Momentenverhaeltnisse} 
several universal moment ratios as obtained 
for the simple contact process (DP), for the pair-contact process
(PCP) without diffusion, and for a branching-annihilating random walk belonging
to the parity-conserving class (PC). The numerical estimates were obtained
at the critical point of finite-size systems by averaging over surviving
quasi-stationary runs and extrapolating $L \to \infty$.
Note that for the contact process,
the order parameter is the particle density, whereas the pair-contact
process with $D=0$, the order parameter is the {\em pair} density. 
While it had been known before that the static
critical exponents of the DP and the PCP agree \cite{Jens93} the agreement
of the moment ratios between the two models provides additional evidence in
favour of the conjecture that the steady-state transitions of the DP and the
PCP are in the same universality class. On the other hand, the result for
the branching-annihilating random walk (BAW) \cite{Zhon95} which is in the 
PC class is clearly different, as expected. 
We are not aware of analogous data
for any other interacting particle model but the building-up of a collection of
data of these amplitudes in several universality classes should be helpful.

\begin{figure}
\centerline{\includegraphics[width=110mm]{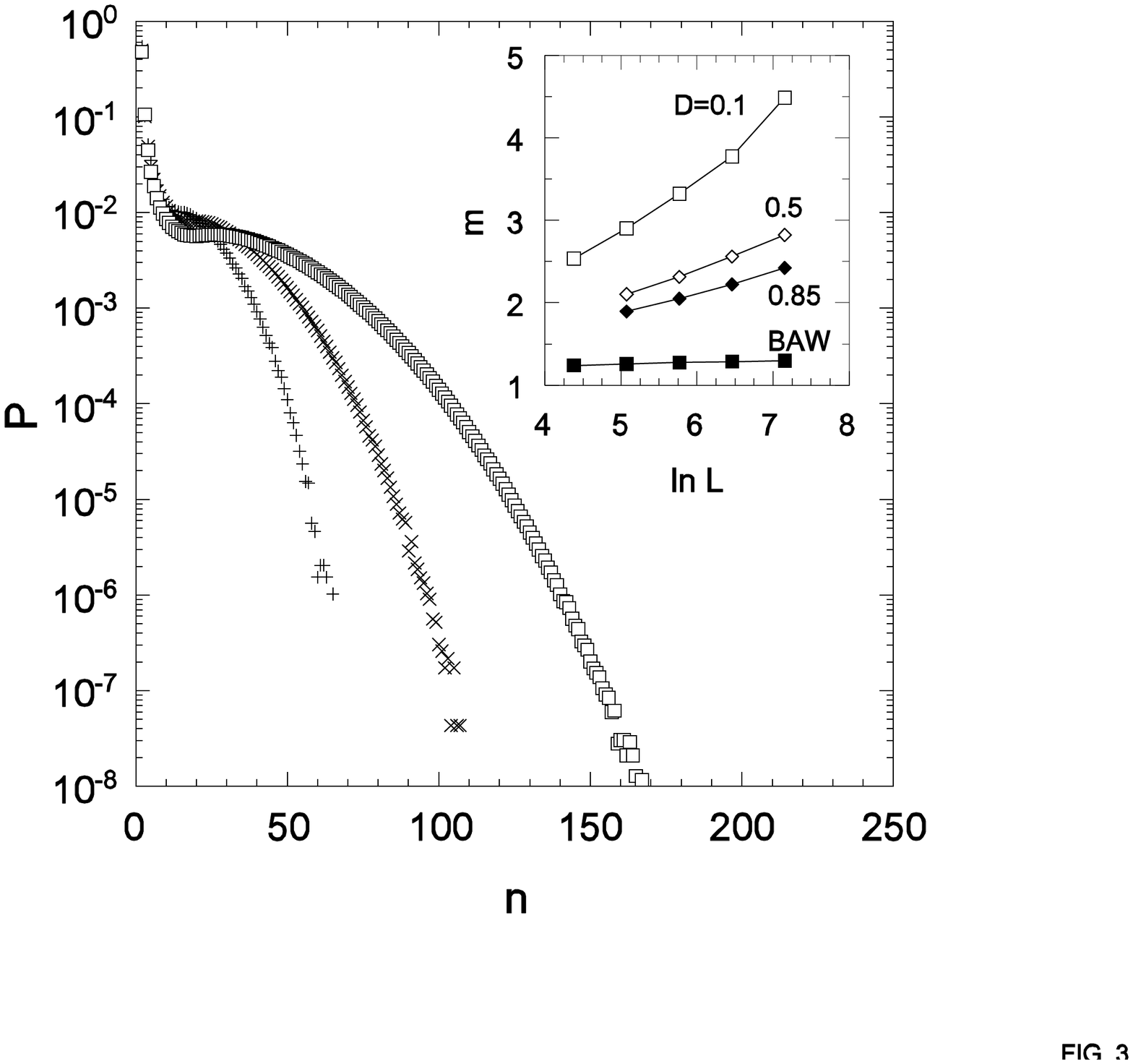}
\hspace{-30mm}\includegraphics[width=118mm]{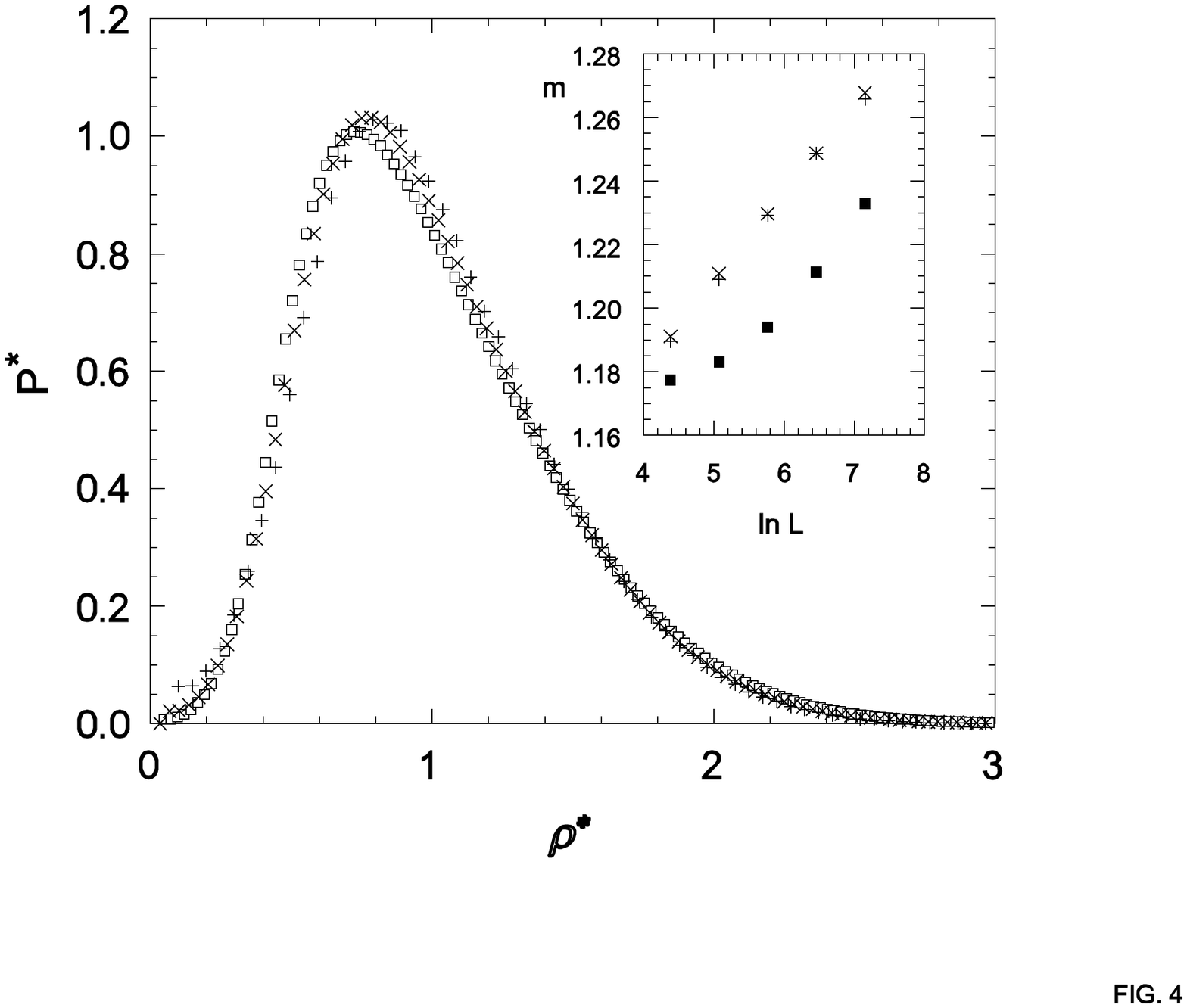}}
\caption[Momentratio]{Left panel: Probability distribution of the 
number of particles $n$ at criticality, for $D=0.1$ and $+$: $L=80$;
$\times$: $L=160$; $\Box$: $L=320$. The inset shows $m=m(\ln L)$ of the
PCPD and also for the BAW.
Right panel: scaling of the reduced probability distribution in the reactive
sector and with the same symbols as before. 
The inset shows $m=m(\ln L)$ in the reactive sector of the PCPD, with
$\blacksquare$: $D=0.1$; $+$: $D=0.5$; $\times$: $D=0.85$. 
After {\protect\cite{Dick02b}}. 
\label{MomentenV}}
\end{figure}

Having established that the method of universal moments is
capable to distinguish between different universality classes, Dickman and
Menezes \cite{Dick02b} tried to apply the moment ratio 
method to the $1D$ PCPD and considered
the universal ratio $m := \langle\rho^2\rangle/\langle\rho\rangle^2=m_2/m_1^2$.
Although several critical exponents could be determined from
finite-size scaling, surprisingly they found for the moment ratio
a logarithmic increase with the system size as $m\sim (\ln L)^{\psi}$ with 
$\psi\gtrsim 1$, see figure~\ref{MomentenV}. No such behaviour is seen for 
the BAW. If this phenomenon persisted in the limit $L\to\infty$,
it would collide with the scaling forms reviewed in section~\ref{sect2}
from which one would deduce a scaling form
\BEQ
P(\rho;L)=\langle\rho\rangle {\cal P}\left(\rho/\langle\rho\rangle\right) \,,
\EEQ
%
%
%
%
where $\cal P$ is a normalized scaling function. Indeed, as can be seen in
figure~\ref{MomentenV}, the particle distribution function does not scale
(a similar result also holds for the pair distribution function). Instead, the
most probable particle number is always 2 and the distribution shows a tail
which grows with $L$ and generates the critical behaviour of the model.

In order to account for this, Dickman and Menezes \cite{Dick02b} argue that 
the PCPD should be characterized by the simultaneous dynamics of isolated 
particles and colliding pairs (as is indeed rendered plausible by illustrations
of the dynamics as in figure~\ref{FIGPCP}) and therefore propose to separate 
the model's states into a `purely diffusive' sector and a `reactive sector' 
which must contain at least a pair of particles. 
The idea is to restrict the analysis to the reactive 
sector by excluding all states without pairs
when taking averages, but without modifying the dynamical
rules (this was also proposed independently in \cite{Noh04}). 
The result is shown in the right panel of figure~\ref{MomentenV}, where
a scaling behaviour in the reduced variables $\rho^*=\rho/\langle\rho\rangle$
and $P^*=\langle\rho\rangle\, P$ is obtained. Now, the form of $P^*$ 
becomes quite similar to the one of the non-diffusive PCP. 
In addition, the ratio $m$ also becomes almost constant and with a value close
to the estimates for DP, see figure~\ref{MomentenV}. 
Still, as already noted in \cite{Dick02b}, neither the
scaling collapse nor the $L$-independence of $m$ is completely perfect and
the origin of these deviations remains to be understood.\footnote{Restricting 
to states with at least 2 pairs does not reduce further the apparent variation 
of $m$ with~$L$.} Although off-critical simulations lead to a clear scaling
collapse and a DP-like exponent $\nu_{\perp}=1.10(1)$ independently of 
$D$ \cite{Dick02b}, other exponents maintain a weak $D$-dependence. When
taking these results at face value, it implies a non-universality 
along the critical line and both the DP and the PC universality classes
are excluded for the $1D$ PCPD \cite{Dick02b}. However, one might wonder 
whether the estimates of the exponents could still be affected by some 
non-resolved correction-to-scaling term which would make the dependence on 
$D$ only apparent. In any case, this study points to the large corrections to 
scaling which are present in the $1D$ PCPD and which render the extraction of
universal parameters very difficult. 

While the above universal moment ratios referred to a spatially infinite
system, we now consider the case of a finite lattice of linear size $L$,
e.g. a hypercube. 
For equilibrium systems, it is well-known that the finite-size scaling
of the correlation length leads to universal amplitudes, as reviewed 
in \cite{Priv93}. For non-equilibrium phase transitions, it has been 
argued that on a finite spatial lattice, the temporal correlation length
$\xi_{\|,i}$ and the spatial correlation length $\xi_{\perp,i}$ should scale
as \cite{Henk01b}
\BEA
\xi_{\|,i}^{-1} = L^{-z} D_0 R_i\left( C_1 (p-p_c)L^{1/\nu_{\perp}}\right)
\nonumber \\
\xi_{\perp,i}^{-1} = L^{-1} S_i\left( C_1 (p-p_c)L^{1/\nu_{\perp}}\right)
\label{Ampxi}
\EEA
%
%
\begin{figure}
\centerline{\includegraphics[width=110mm]{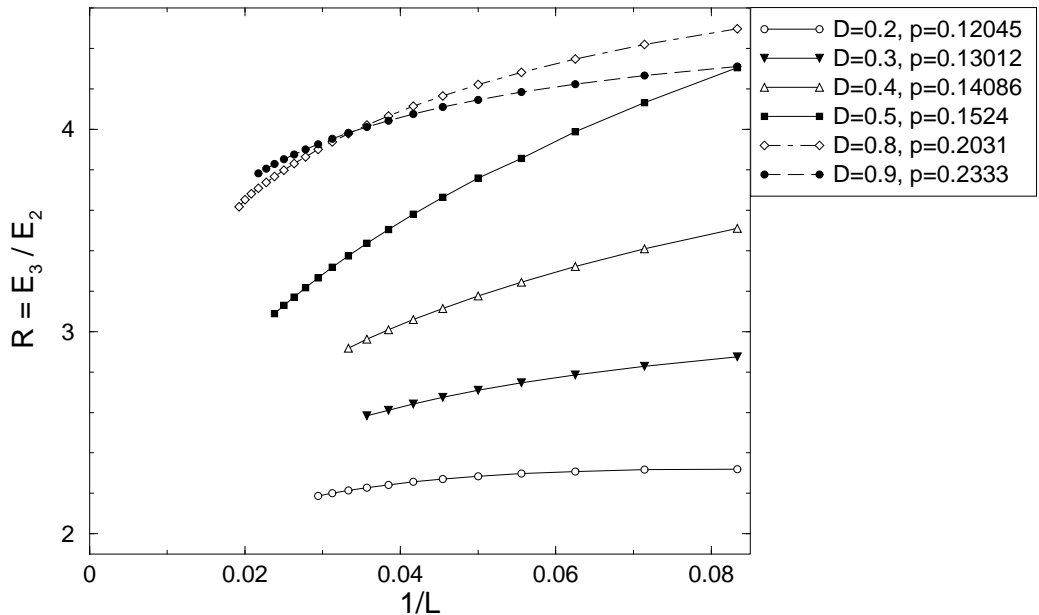}}
\caption[Amplitude ratios]{Universal ratio of finite-size scaling amplitudes
of the relaxation times along the critical line in the $1D$ PCPD, for free
boundary conditions. The values of $p_c(D)$ are from 
{\protect\cite{GrasPrivat}}. After {\protect\cite{Henk01b}}. 
\label{AmplitudenV}}
\end{figure}
%
%
where (for given boundary conditions) $R_i$ and $S_i$ are {\em universal} 
scaling functions, the index $i$ refers
to the observable $\mathfrak{O}_i$ of which the correlators are studied 
and the entire non-universality can be absorbed into the non-universal 
metric factors $C_1$ and $D_0$. In particular, the existence of the above 
scaling forms can be {\em derived} for systems in the directed percolation 
universality class \cite{Henk01b}. For the convenience of the reader, the 
argument will be reproduced in appendix~C.
 Consequently, the finite-size scaling 
amplitudes of the spatial correlation lengths $\xi_{\perp}^{-1}$ are universal,
as are {\em ratios} of temporal correlation lengths 
$\xi_{\|,i}^{-1}/\xi_{\|,j}^{-1}=E_i/E_j$ which in turn are easily calculated
from the spectrum of the quantum hamiltonian $H$, as discussed above in
section~4.1. Furthermore, from the scaling forms derived in
appendix~C it follows that the universality of the critical finite-size 
scaling amplitudes of $\xi_{\perp,i}$ is equivalent to the 
universality \cite{Henk01b} of the moments
\BEQ \label{Ampr}
\left\langle r_{\perp}^n \right\rangle = 
\frac{\int_{\Omega(L)}\!\D^d\vec{r}_{\perp}
\int_0^{\infty}\!\D r_{\|}\,r_{\perp}^n 
G(\vec{r}_{\perp},r_{\|};L/\xi_{\perp})}{\int_{\Omega(L)}\!\D^d\vec{r}_{\perp}
\int_0^{\infty}\!\D r_{\|}\,
G(\vec{r}_{\perp},r_{\|};L/\xi_{\perp})} 
= L^n\, \Xi(L/\xi_{\perp}) \,,
\EEQ
of the pair connectivity $G(\vec{r}_{\perp},r_{\|})$ which is the probability
that the sites $(\vec{0},0)$ and $(\vec{r}_{\perp},r_{\|})$ are connected by
a direct path. Here $\Omega(L)$ is a $d$-dimensional hypercube of linear
extent $L$ and $\Xi$ is a universal function the argument of which does not 
contain any non-universal metric factor. These universality 
statements have been confirmed in several exactly solvable models
\cite{Henk01a,Henk01b}. In figure~\ref{AmplitudenV} we illustrate the 
universality of the amplitude ratio $R=E_3/E_2$ along the critical line
in the $1D$ PCPD, as obtained from the DMRG. 
As observed before for the exponents, there appear to be
some finite-size corrections which are sufficiently irregular such that the 
powerful sequence-extrapolation methods \cite{Henk88,Barb82} cannot be used. 
Still, inspection shows that for the smaller values of $D$ (i.e. up to
$D\lesssim 0.5$) the data seem to converge towards a limit 
$R_{\infty}\approx 2$ and its independence of $D$ would confirm universality. 
On the other hand, for values of $D\gtrsim 0.8$, there appears to arise
a large transient region and only for values of $L$ larger than $\approx 50$
sites a convergence might slowly set in. This is a cautionary example
illustrating again that the truly asymptotic regime in the PCPD might set
in very late and sometimes could be beyond the reach of currently available 
numerical methods. 

At present, no finite-size data for the amplitude ratios (\ref{Ampxi}) or
(\ref{Ampr}) seem to be available for free boundary conditions in any 
nonequilibrium universality class. In this context we recall 
the well-known observation from critical 
equilibrium systems that universal amplitude ratios tend to vary considerably 
more between different universality classes than usually do critical exponents,
see \cite{Priv93}. Finding reference values for these amplitudes in distinct
non-equilibrium systems should be helpful. 

\subsection{Extrapolation techniques}
\label{Extrapolation}

In numerical simulations of the PCPD there are two sources of uncertainty 
which cannot be separated easily. On the one hand, the critical PCPD shows 
unusually strong {\em intrinsic corrections} to scaling, on the other hand 
a possible error in the estimation of the critical point leads to 
additional {\em off-critical deviations} with an unknown functional form. 
Most numerical approaches assume that the first source of uncertainty can 
be neglected after sufficiently long simulation time. The critical point 
is then estimated by postulating power-law behaviour over the last few decades 
in time, indicated by a straight line in a log-log plot or saturating 
effective exponents. In the PCPD, however, the main risk lies in the 
possibility that the intrinsic corrections to scaling {\em at} criticality 
may still be present even after $10^6\ldots 10^8$ time steps. 
Thus, looking for a power-law over the last few decades of time, 
one may be tempted to `compensate' the intrinsic scaling corrections 
by off-critical deviations in opposite direction, leading to a slightly 
biased estimate of the critical point and thereby to considerable systematic 
errors in the estimated critical exponents. 

In order to solve this problem the two sources of corrections -- 
intrinsic scaling corrections and off-critical deviations -- 
have to be separated. One may try to do this by assuming a certain functional 
form for one of them and performing appropriate fits. The first attempt 
in this direction was made by \'Odor~\cite{Odor02c}, who postulates 
logarithmic corrections to scaling at criticality. For example, in the case of 
a temporally decaying particle density, where we expect an asymptotic 
power law $\rho(t)\sim t^{-\delta}$, he assumes the intrinsic scaling 
correction {\em at criticality} to be of the form
\begin{equation}
\rho(t) = \left[ \frac{a+b \ln t}{t} \right] ^\delta \ ,
\end{equation}
where $a$ and $b$ are fit parameters. Using this technique \'Odor 
obtained $\delta=0.21(1)$ for various values of the diffusion rate 
between 0.05 and 0.7, favouring the scenario of a single independent 
universality class. \'Odor also estimated the critical exponents in a 
two-dimensional PCPD, confirming the mean-field prediction in $d \geq 2$
dimensions.

\begin{figure}
\centerline{\includegraphics[width=90mm]{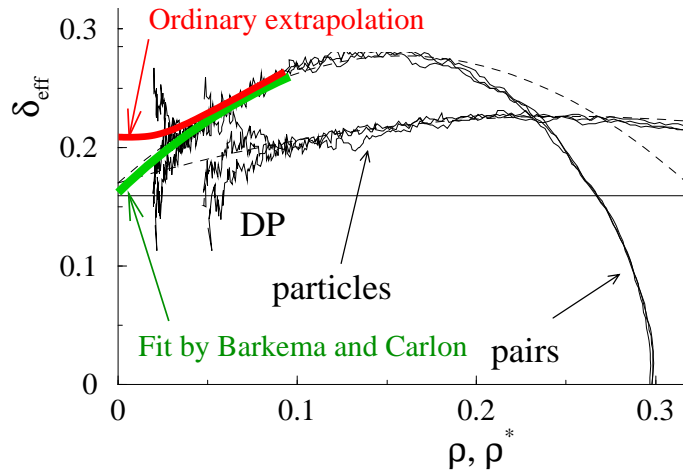}}
\caption{\label{figbarkema}
Extrapolation technique suggested by Barkema and Carlon 
(after \cite{Bark03}). Assuming algebraic corrections  
of the form~(\ref{BarkemaCorrections}) with $\gamma=\delta$ 
the curve of the effective exponent $\delta_{\rm eff}(t)$ should intersect 
with the vertical axis as a straight line (lower bold line). Ordinary 
extrapolations such as in Ref.~\cite{Kock02}, assuming that the scaling 
regime is actually reached within the simulation time, would correspond 
to the upper bold line, which becomes horizontal as $\rho(t)\to 0$.
}
\end{figure}

Another extrapolation method was tried by Barkema and 
Carlon~\cite{Bark03} in analysing high-quality data of a multispin Monte
Carlo simulation. Unlike \'Odor, they postulate algebraic scaling 
corrections of the form
\begin{equation}
\label{BarkemaCorrections}
\rho(t) = (1+at^{-\gamma})\, t^{-\delta} \ ,
\end{equation} 
where $a$ and $\gamma$ have to be fitted appropriately. 
Although one would naively expect the exponent of the correction 
$\gamma$ to be equal to $1$, Barkema and Carlon present numerical 
evidence suggesting that $\gamma=\delta=\beta/\nuperp$. Based on this 
conjecture they propose to plot local slopes versus the density of 
particles. Using this representation the curve of local effective 
exponents should intersect with the vertical axis at $\rho=0$ as a 
straight line, as sketched in figure~\ref{figbarkema}. Applying this 
extrapolation method Barkema and Carlon argue that the critical 
exponents tend to those of DP, in agreement with the conclusions
offered in \cite{Hinr02a} (they only quote the value of $\delta$ averaged over
several values of $D$ which is the one listed in table~\ref{ExpTable}).
%
%
%
%

\subsection{Restricted PCPD with a soft constraint}
\label{Kockelkoren}

All variants of the PCPD discussed so far are restricted by hard-core exclusion,
i.e. each site can be either empty or occupied by a single particle. 
Although these models are easy to implement numerically, they involve 
three-site interactions with some ambiguity concerning the spatial arrangement.
In order to circumvent
this problem, Kockelkoren and Chat\'e~\cite{Kock02} proposed a surprisingly 
simple `bosonic' model in which a soft constraint prevents the particle density 
in the active phase from diverging. 
The scaling arguments of the previous sections are expected to go through and 
models with a soft constraint are expected to 
exhibit the same type of critical behaviour as the ordinary PCPD with 
hard-core exclusion, see also section~6.

Their model is defined as follows. Particles of a single species $A$ evolve by 
synchronous updates in two sub-steps. Each site can be occupied by arbitrarily 
many particles. First the particles diffuse, i.e. all particles hop 
independently to a randomly chosen nearest neighbour. Then the pairs of 
particles react locally, either producing offspring or annihilating each other. 
To this end the population
of $n$ particles at a given site is divided into $\lfloor n/2\rfloor$ pairs,
which branch independently with probability $p^{\lfloor n/2\rfloor}$ and 
annihilate otherwise. As a result, if $n$ is very large, the fission process is
exponentially suppressed, introducing effectively a soft constraint.

\begin{figure}
\centerline{\includegraphics[width=100mm]{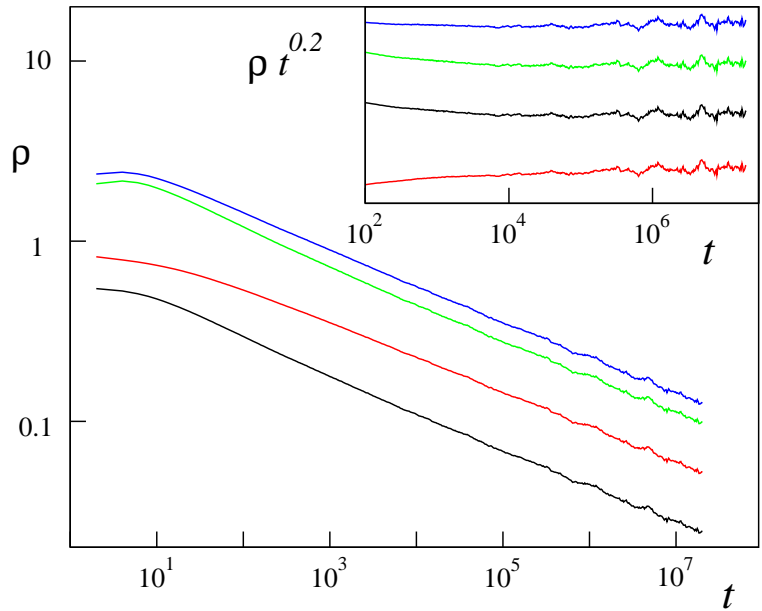}}
\caption[Kockelkoren]{
Soft-constraint model defined in {\protect\cite{Kock02}} at criticality.
{}From top to bottom are shown the decay of the particle density,
the density of particles without solitary particles, the fraction 
of occupied sites and the fraction of occupied sites with at least two 
particles. A clean scaling behaviour is observed.
After {\protect\cite{Kock02}}. 
\label{FigKockelkoren}}
\end{figure}

Unlike restricted variants of the PCPD, the model introduced by Kockelkoren and
Chat{\'e} exhibits a surprisingly clean scaling behaviour, even when different
order parameters are used (see figure~\ref{FigKockelkoren}). 
Since all reactions
take place on single sites, the model seems to reach the scaling regime much
faster than ordinary models with hard-core exclusion. 
These results led the authors of \cite{Kock02} to the
conclusion that the PCPD represents a single universality class different from
DP and PC, characterized by the critical exponent 
$\delta \approx 0.20$.

\subsection{Surface critical behaviour}
\label{Surfaces}

While most of the attempts in understanding the critical behaviour of the
PCPD concentrated on bulk quantities, some information of surface critical
exponents is also available. It is a well-established fact, that in a
semi-infinite system the critical behaviour near to the surface is in general
different from the one deep in the bulk, see \cite{Froj01} for a recent
review. For example, in the steady-state
the order-parameter density should scale near criticality as
$\rho_{\rm b}\sim (p-p_c)^{\beta}$ in the bulk but near the surface one
expects $\rho_{\rm surf}\sim (p-p_c)^{\beta_1}$ where $\beta_1$ is a surface
critical exponent. The value of $\beta_1$ may further depend on the
boundary conditions. Here we shall consider
\begin{enumerate}
\item {\em free} boundary conditions, i.e. particles cannot cross the boundary.
\item {\em absorbing} boundary conditions, i.e. particles {\em at} the 
boundary may leave the system with a rate $D$. In other words, the boundary 
reaction $A\to\emptyset$ is added. 
\end{enumerate}
At criticality, one expects the finite-size scaling behaviour
$\rho_{\rm surf}\sim L^{-\beta_1/\nu_{\perp}}$ and techniques to calculate
the boundary density are available \cite{Carl99}. For the PCPD, Barkema
and Carlon \cite{Bark03} have obtained estimates for the exponent ratio
$\beta_1/\nu_{\perp}$ and we list their results in table~\ref{oberfl} together
with results for the DP and PC universality classes for comparison. In the
PCPD, estimates only converge for $D\lesssim 0.5$. 

\begin{table}
\footnotesize
\begin{center}
\begin{tabular}{|c|ll|l|c|} \hline
model & \multicolumn{1}{c}{free} & \multicolumn{1}{c|}{absorbing} & 
\multicolumn{1}{c|}{method} & Ref. \\ \hline
PCPD & $0.72(1)$   & $1.11$    & DMRG, $D\lesssim0.5$ 
                                                    & \cite{Bark03}\\\hline
DP   & $0.664(7)$  & --        & Monte Carlo        & \cite{Froj01}\\
     & $0.667(2)$  & --        & DMRG               & \cite{Carl99}\\
     & $0.6690(1)$ & --        & series             & \cite{Essa96}\\\hline
PC   & $0.73(1)$   & $1.11(1)$ & Monte Carlo        & \cite{Froj01}\\
     & $0.720(2)$  & $1.10(1)$ & DMRG               & \cite{Bark03}\\\hline
\end{tabular} 
\caption{\label{oberfl} Values of the surfrace exponent $\beta_1/\nu_{\perp}$
according to different methods for free and absorbing boundary conditions in
some $1D$ models.}
\end{center}
\end{table} 

Taken at face value, the estimates of $\beta_1/\nu_{\perp}$ obtained for the 
PCPD appear actually to be compatible with those of the PC class but Barkema 
and Carlon carefully point out that their final estimates still depend on $D$, 
in particular for $D$ large, and therefore may not yet give the correct 
asymptotic values. 

\section{Related models}

\subsection{Cyclically coupled spreading and annihilation}
\label{CycSect}
   
%
%
\begin{figure}
\centerline{\includegraphics[width=90mm]{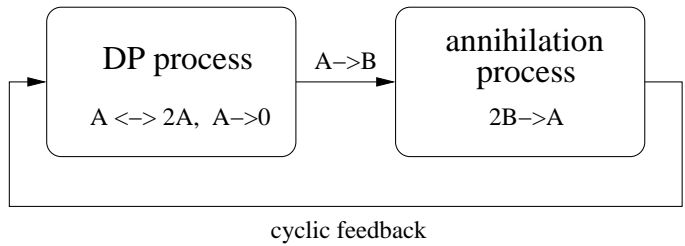}}
\caption{\label{FIGCYCLIC}
PCPD interpreted as a cyclically coupled DP and annihilation process.
}
\end{figure}
%
%

In order to characterize the essential features of the PCPD from a 
different perspective, one of us (H.H.) suggested to interpret the PCPD as 
a cyclically coupled DP and annihilation process~\cite{Hinr01b}. 
The idea is to separate the dynamics of pairs and solitary particles, 
associating them with two species of particles $A$ and $B$. The $A$'s, 
which stand for the pairs in the original PCPD, perform an ordinary DP process 
while the $B$'s represent solitary particles that are subjected to an 
annihilating random walk. Both subsystems are  cyclically coupled by particle 
transmutation, as sketched in figure~\ref{FIGCYCLIC}. 

Such a cyclically 
coupled process was realized in \cite{Hinr01b} as a three-state model with 
random-sequential updates. Choosing particular rates it was shown that the 
process exhibits a phase transition with the same phenomenology as in the PCPD.
Moreover, the values of the effective critical exponents
\begin{equation}
\delta=0.215(15), \quad z=1.75(5), \quad\nupar=1.8(1) 
\end{equation} 
were found to be in fair agreement with other estimates for the PCPD. 
Recent simulations for various values of $D$ 
between $0.1$ and $1$ suggest that the exponents should be independent 
of the diffusion constant~\cite{ParkPrivat}. These results indicate 
that cyclically coupled processes following the reaction-diffusion scheme 
in figure~\ref{FIGCYCLIC} display the same type of transition as the PCPD.

\begin{figure}
\centerline{\includegraphics[width=160mm]{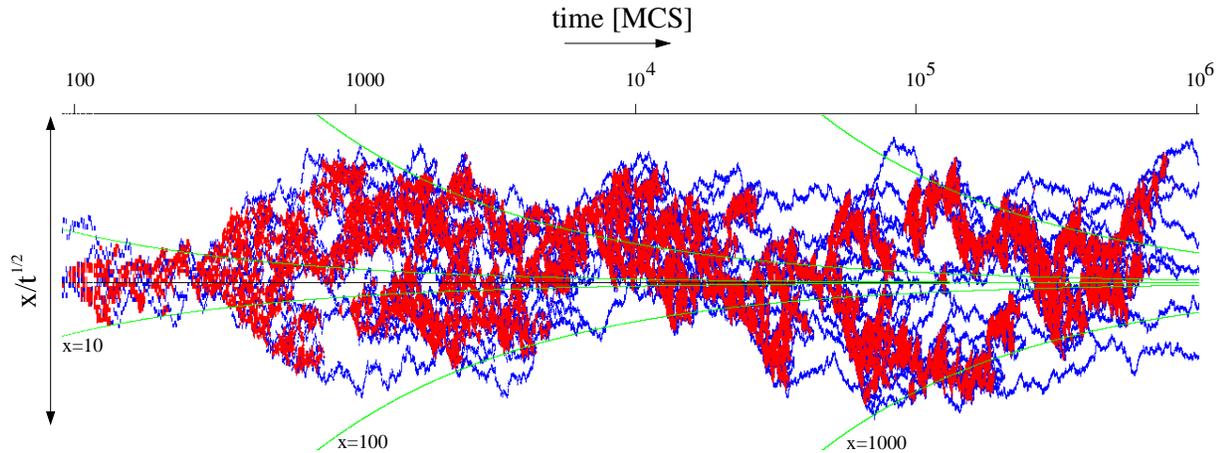}}
\caption{\label{FIGAB} (colour online)
Typical spatio-temporal evolution of a binary spreading process starting
from an initial seed. Particles of species A and B are represented as red
and blue pixels, respectively. Plotting $x/L^{1/2}$ versus $\log_{10} t$ 
the figure covers four decades in time. As can be seen, patches of high 
activity (red) are connected by lines of diffusing $B$-particles (blue) 
on all scales up to $10^6$ time steps. 
}
\end{figure}

The separation into subprocesses 
illustrates an important feature of the PCPD, namely, the existence of 
two modes of spreading. As illustrated in figure~\ref{FIGAB}, 
a typical spatio-temporal cluster is 
characterized by the interplay of a high-density mode dominated by 
self-reproducing and decaying $A$-particles, 
and a low-density mode of solitary diffusing $B$-particles. 
A similar interplay of high- and low-density patches can be observed in 
the critical PCPD, see figure~\ref{FIGPCP}.
%
%
%
%

Concerning the structure of a cluster
a fundamental problem arises precisely at this point: As can be seen in 
figure~\ref{FIGAB}, even after $10^6$ time steps the solitary particles 
perform simple random walks over large distances. However, such a 
random walk is always characterized by the dynamic exponent $z=2$, 
while all known simulations clearly indicate that the process as a whole 
spreads {\em superdiffusively}, i.e. $z<2$. 
Therefore, the effective diffusion constant for solitary particles has 
to vary slightly under rescaling, meaning that a cluster such as in 
figure~\ref{FIGAB} cannot be scale-invariant. Therefore it seems that 
the process is still far away from the asymptotic scaling regime, even 
after $10^6$ time steps.

\subsection{Interpolating between DP and PCPD}
   
A different explanation of the apparent non-universality in the PCPD 
was suggested by Noh and Park~\cite{Noh04}, who claim that the violations of 
scaling can be traced back to a long-term memory effect
mediated by the diffusive background of single particles. To show this, they 
introduce a generalized version of the PCPD, as follows. Each site of the
lattice can be either empty or occupied by a single particle and the
following reactions are admitted
\begin{eqnarray}
AA \to \emptyset\emptyset && \mbox{\rm with rate $(1-D)p$} \nonumber\\
AA\emptyset,\emptyset AA \to AAA ~~~&& \mbox{\rm with rate $(1-D)(1-p)/2$}
\\ \nonumber
A\emptyset\emptyset\to\emptyset A\emptyset && \mbox{\rm with rate $D$} \\
A\emptyset A\to\emptyset AA && \mbox{\rm with rate $Dr$}\\ \nonumber
A\emptyset A\to\emptyset\emptyset\emptyset && \mbox{\rm with rate $D(1-r)$}
\nonumber
\end{eqnarray}
which allows one to interpolate between the DP fixed point at $r=0$ and the 
critical behaviour of the PCPD at $r=1$. 
In this model the branching probability depends on whether a
collision is caused by a previous branching process or by diffusion.
In addition, one restricts the calculation of averages to the sector where
at least one pair (rather than two isolated particles) is present, in analogy
with \cite{Dick02b}. Since $r=1$ in this model is not a special point, 
one may hope that corrections to scaling are more easy to control for $0<r<1$ 
and that the results could be extended to $r=1$ at the end. 

Dynamic and
static exponents are found from simulations at several values of
$r$ and some of the results of \cite{Noh04} are listed in table~\ref{GPCPD}.
As a consistency check, the exponent values are used to test the 
hyperscaling relation eq.~(\ref{HYPER}) and a full agreement is observed. 

\begin{table}
\footnotesize
\begin{center}
\begin{tabular}{|l|l|lllll|} \hline
\multicolumn{1}{|c|}{$r$} & \multicolumn{1}{c|}{$p_c$} 
& \multicolumn{1}{c}{$z$} 
& \multicolumn{1}{c}{$\eta$} & \multicolumn{1}{c}{$\delta'$} 
& \multicolumn{1}{c}{$\delta=\beta/\nu_{\|}$} 
& \multicolumn{1}{c|}{$\nu_{\perp}$} \\ \hline
0.00 & 0.04687(2) & 1.58(1) & 0.314(6) & 0.160(5) & 0.159(1) & 1.10(1)  \\
0.25 & 0.05505(5) & 1.62(3) & 0.29(1)  & 0.15(1)  & 0.173(5) & 1.10(3)  \\
0.50 & 0.06636(4) & 1.67(3) & 0.26(1)  & 0.14(1)  & 0.197(3) & 1.10(3)  \\
0.75 & 0.08315(5) & 1.75(5) & 0.20(2)  & 0.13(2)  & 0.230(5) & 1.17(5)  \\
1.00 & 0.1112(1)  & 1.7(1)  & 0.18(5)  & 0.09(2)  & 0.27(4)  & 1.30(10) \\\hline
\end{tabular} 
\caption{\label{GPCPD} Critical parameters of the generalized PCPD as defined
in {\protect\cite{Noh04}} for $D=0.1$ as a function of $r$.}
\end{center}
\end{table} 

Comparison with table~\ref{ExpTable} shows that for $r=0$, these results are
in excellent agreement with the DP values while for $r=1$, they fall within
the range of values of the PCPD determined with different methods. 
In contrast to the limit
$D\to 0$, where the exponents change abruptly, the exponents seem to vary
smoothly when $r$ is taken to zero. Based on these numerical observations
Noh and Park \cite{Noh04} conclude that the transition in the PCPD may 
correspond to a line of fixed points with continuously varying exponents. 
As a possible explanation they suggest to consider the long-term memory imposed 
by solitary diffusing particles as a marginal perturbation of the underlying 
field-theory.

\subsection{Parity-conserving PCPD}

As the PC class compared to DP exemplifies that an additional 
parity-conservation law may change the universality class of an 
absorbing phase transition
the question arises whether the same happens in binary spreading processes.
To answer this question in the context of the PCPD parity-conserving process
\begin{equation}
2A \to 4A \,, \qquad 2A \to \emptyset \,.
\end{equation}
was studied \cite{Park01}. Surprisingly it turns out that
parity-conservation does not change the nature of the phase transition, i.e.,
one observes the same phenomenological behaviour as in the PCPD with comparable
effective critical exponents $\delta \simeq 0.23(1)$ and $z=1.80(5)$.

To understand this observation, we note that there is another
well-known example where parity conservation is irrelevant,
namely, the annihilation process $2A \rightarrow 0$  as compared to
the coagulation process $2A \rightarrow A$, which are known to belong to 
the same universality class. This is due to the fact 
that the even sector and the odd sector in the parity-conserving
process $2A \rightarrow 0$ are essentially equivalent since
in both of them the particle density decays algebraically until 
the system is trapped in an absorbing state (namely, the empty lattice 
or a state with a single diffusing particle). The present model 
is similar in so far as both sectors have an absorbing state.
Contrarily, in the PC class only one of the sectors has an absorbing
state, leading the authors of~\cite{Park01} to the conclusion that
parity conservation changes universality whenever one of the two sectors
has no absorbing state.

\subsection{Multicomponent binary spreading processes}

In order to search for further possible generalizations, \'Odor~\cite{Odor02b} 
investigated different variants of binary production-annihilation processes 
with several species of particles, in particular the reaction-diffusion 
processes
\begin{itemize}
\item[(a)] \ $\quad AB\to ABA, \quad 2A\to\emptyset, 
\quad BA\to BAB, \quad 2B\to\emptyset$.
\item[(b)] \ $\quad 2A\to 2AB, \quad \hspace{3mm} 2A\to\emptyset, 
\quad 2B\to 2BA, \hspace{2mm} \quad 2B \to\emptyset$
\end{itemize}
The process (a) seems to exhibit the same type of critical behaviour as the 
PCPD, wile in the process (b) the phase transition takes place at zero 
branching rate, where the exponents are different, e.g. $\beta=2$. 
However, it would be misleading to conclude
that the process (b) indeed represented a different universality class, 
rather the model
is designed in a way that the nontrivial transition, which is possibly 
belonging to the same class as the PCPD, is shifted to an inaccessible 
region of the parameter space. Detecting a transition at zero branching 
rate does not necessarily mean that all realizations of the same 
reaction-diffusion scheme will yield such a trivial transition.

\subsection{Roughening transition driven by the PCPD}

As demonstrated in~\cite{Hinr02b} the PCPD may also play a role 
in a special class of models for 
interface growth which exhibit a roughening transition. The idea follows an 
earlier work by Alon \etal~\cite{Alon96,Alon98}, where a deposition-evaporation 
model with a DP-related roughening transition was introduced. Later it was 
shown that by depositing dimers instead of monomers one obtains a different 
type of roughening transition which is related to the PC 
class~\cite{Hinr99a,Hinr99b}. 

As a key property of all these models, desorption is only allowed at the 
{\em edges} but not from the middle of deposited plateaus. Therefore the actual 
bottom layer of the interface, once it has been completed, cannot evaporate 
again which leads effectively to an absorbing phase transition at the bottom 
layer. Using this interpretation the pinned phase, where the interface is 
smooth, corresponds to the active phase of the underlying spreading process. 
However, if the growth rate is increased above a certain critical threshold the 
interface eventually detaches, meaning that the bottom layer enters an 
`absorbing' state.

\begin{figure}
\flushright{\includegraphics[width=135mm]{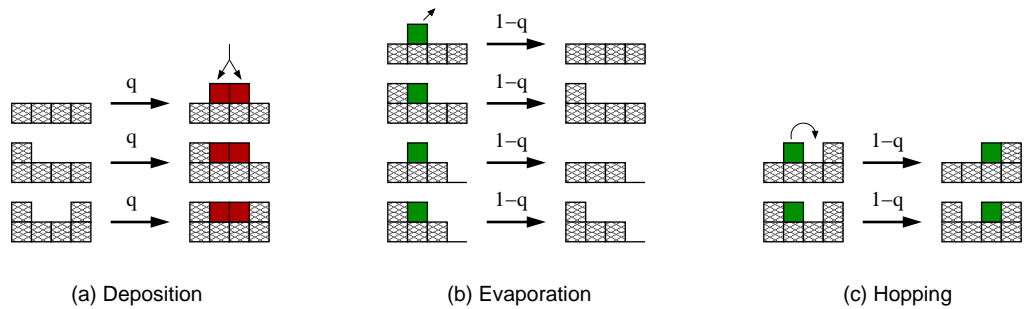}}
\caption{\label{figgrowthmodel}
Dynamic rules of the PCPD-driven growth process: {\bf (a)} Deposition of a 
dimer at the sites $i$ and $i+1$. {\bf (b)} Evaporation of a monomer from 
the right edge of a terrace at site $i$. {\bf (c)}~Hopping of a monomer 
between two adjacent islands, moving the hole between the islands in 
opposite direction. In all cases the spatially reflected rules 
(not shown here) have to be included as well.
}
\end{figure}

In order to study a PCPD-driven roughening transition one has to introduce 
appropriate dynamic rules that mimic a binary spreading process at the 
bottom layer. These rules involve three different physical processes, 
namely, deposition of dimers, surface diffusion, and evaporation of 
diffusing monomers (see figure\ref{figgrowthmodel}). The dimers are deposited 
horizontally on pairs of sites at equal height, leading to the formation of 
islands. These islands are stable in the interior but unstable at 
the edges, where monomers are released at a certain rate. The released monomers 
diffuse on the surface until they either attach to another island or 
evaporate back into the gas phase. In the limit of a very high 
evaporation-rate a released monomer is most likely to evaporate unless 
it immediately attaches to an {\em adjacent} island at the next site, 
effectively moving the hole between the two islands in the opposite direction. 

In order to explain how the growth model is related to the PCPD, let us 
consider the dynamic processes at the bottom layer (the spontaneously selected 
lowest height level of the interface). Interpreting sites at the bottom layer 
as particles $A$ of a spreading process, the dynamic rules listed in 
figure~\ref{figgrowthmodel} can be associated with certain reactions of 
the particles. For example, the deposition of a dimer corresponds to pairwise 
annihilation $2A\to\emptyset$, while evaporation of a monomer can be viewed 
as the creation of a particle $A$. However, in the present model atoms can 
only evaporate at the edge of a terrace followed by two bottom layer sites, 
hence rule {\bf (b)} in figure~\ref{figgrowthmodel} has to be interpreted as a 
fission process $2A\to 3A$. Otherwise, if there is only one such bottom 
layer site next to the edge, rule~{\bf (c)} applies which corresponds to 
a random walk of a single particle $A$. Thus the processes at the bottom 
layer resemble the dynamic rules of the PCPD. Clearly, this correspondence is 
not rigorous, especially because of second-layer nucleations. However, as in 
the case of DP- and PC-related growth models, the correspondence is expected to 
be valid asymptotically.

In the case of DP-related growth processes, the behaviour at the first few 
layers above the bottom layer can be described by a unidirectionally coupled 
hierarchy of DP processes~\cite{Gold99}. Similarly the dynamics at higher 
levels in the growth process in figure~\ref{figgrowthmodel} is expected to be 
related by a unidirectionally coupled sequence of PCPD's, which has not been 
investigated so far.

\section{Field-theoretic approaches}\label{FTSec}

Since the numerical simulations as a whole remain inconclusive, it is 
important to find and analyze a suitable field-theory that 
describes the phase transition in the PCPD. So far there are two important 
contributions in this direction. Some time ago Howard and 
T{\"a}uber~\cite{Howa97} proposed a field-theory for the unrestricted PCPD,
while recent attempts to devise a field-theory for the restricted case
turned out to be unsuccessful~\cite{DJTW}.
Moreover, in a recent paper the pair correlation function 
in the unrestricted PCPD has been derived exactly by Paessens and 
Sch\"utz~\cite{Paes03b}, leading to non-trivial and surprising results. 
The purpose of this section is to summarize the present state of knowledge.

\subsection{Field-theory of the unrestricted case}

Consider the unrestricted PCPD which is defined by the 
reaction-diffusion scheme
\begin{equation}
2A \stackrel{\mu}{\longrightarrow} \emptyset\,,\quad
2A \stackrel{\nu}{\longrightarrow} A\,,\quad
2A \stackrel{\sigma}{\longrightarrow} 3A \,,
\end{equation}
where $\mu,\nu$, and $\sigma$ are the rates for annihilation, coagulation, and 
fission, respectively. In the continuum limit (see e.g.~\cite{Lee94}) the 
classical master equation of this process corresponds to the field-theoretic 
action
\begin{equation}
\label{UnshiftedUnrestricted1}
\!\!\!\!\!\!\!\!\!\!S = \int \!\D^d\vec{x} \int \!\D t\, 
\biggl[
\hat{\phi}(\partial_t-D\nabla^2)\phi - \mu (1-\hat{\phi}^2)\phi^2 
- \nu (1-\hat{\phi})\hat{\phi}\phi^2 + \sigma(1-\hat{\phi}) \hat{\phi}^2\phi^2 
\biggr]
\end{equation}
where $D$ is the diffusion rate while $\hat{\phi}(x,t)$ and $\phi(x,t)$ can be 
thought of as particle creation and annihilation operators. Shifting the 
response field by $\hat{\phi}=1+\bar{\phi}$ the action can also be written as
\begin{equation}
\label{ShiftedUnrestricted}
\!\!\!\!\!\!\!\!\!S = \int \!\D^d\vec{x} \int \!\D t \,
\biggl[
\bar{\phi}(\partial_t-D\nabla^2)\phi + (2\mu+\nu-\sigma)\bar{\phi}\phi^2 
+ (\mu+\nu-2\sigma)\bar{\phi}^2\phi^2 - \sigma\bar{\phi}^3\phi^2
\biggr]
\end{equation}
Since higher-order terms will be generated under renormalization group
(RG) transformations, it is 
convenient to write the action in the general form
\begin{equation}
S = \int \!\D^d\vec{x} \int \!\D t \, 
\biggl[
\bar{\phi} (\partial_t - D \nabla^2) \phi +
\sum_{p=1}^\infty \sum_{q=2}^\infty \Lambda_{p,q}\, \bar{\phi}^p \phi^q 
\biggr]
\end{equation}
taking all possible reactions into account which are at least binary, i.e. 
quadratic in $\phi$. In this notation the bare coupling constants of the 
unrestricted PCPD are given by
\begin{equation}
\label{BareCouplings1}
\Lambda_{1,2}=2\mu+\nu-\sigma\,, \qquad
\Lambda_{2,2}= \mu+\nu-2\sigma\,, \qquad
\Lambda_{3,2}=-\sigma
\end{equation}
while all other {\em bare} coupling constants (but not necessarily the 
renormalized ones) vanish. Regarding the first two terms 
$\bar{\phi}\phi^2$ and $\bar{\phi}^2\phi^2$ a simple 
power-counting analysis yields the upper critical dimension
\begin{equation}
d_c=2
\end{equation}
at which the dynamical critical exponent $z=2$. In the annihilation phase the 
coupling constants and the fields carry the dimensions
\begin{equation}
[\Lambda_{p,q}]=\kappa^{2-(q-1)d}\,, \qquad
[\phi]=\kappa^d\,,\quad
[\bar{\phi}] = 1
\end{equation}
where $\kappa$ denotes an arbitrary momentum scale. Hence for given $q$ all 
contributions with $p$ running from $1$ to $\infty$ are {\em equally relevant}
and the usual renormalization group scheme 
in terms of a finite series of Feynman 
diagrams cannot be applied. At the critical point, however, the dimensions of
the fields are given by 
$[\phi]=[\bar{\phi}] = \kappa^{d/2}$~\cite{JanssenPrivate03}.

In Ref.~\cite{Howa97} the field-theory was found to be 
non-renormalizable so that no quantitative information on critical exponents 
can be obtained. Nevertheless it is possible to compute the 
critical point of the unrestricted PCPD exactly. To this end we note that 
the transition expressed in bare parameters takes place at 
$2 \mu + \nu = \sigma$, where the cubic term $\bar{\phi}\phi^2$ in the shifted 
action vanishes~\cite{Howa97}. Since there is no way to generate this 
contribution under RG transformations to {\em all} orders, fluctuation 
effects do not influence the location of the transition, hence
\begin{equation}
2 \mu + \nu = \sigma
\end{equation}
determines the critical line to all orders of perturbation theory. In the 
inactive phase $2 \mu + \nu < \sigma$ annihilation and/or coagulation 
eventually dominate, leading to an asymptotic decay as $t^{-d/2}$ for $d<2$, 
while in the supercritical regime  $2 \mu + \nu > \sigma$ the particle 
density grows rapidly and diverges within finite time. Consequently, there 
is no stationary state in the supercritical regime. 

\subsection{Attempts towards a field-theory of the restricted PCPD}

Recently van Wijland, T{\"a}uber, and Deloubri\`ere proposed and 
studied a field-theory for the restricted case. Their results, however, 
turned out to be unphysical~\cite{DJTW}, suggesting that a different 
field-theoretic approach is needed.

Starting point of the theory by Wijland {\it et al} was to extend the 
field-theoretic action~(\ref{UnshiftedUnrestricted1}) by an additional 
mechanism which prevents the field $\phi$ from diverging in the 
supercritical regime. This mechanism may be implemented as a soft 
constraint by adding a reaction of the form $3A \to pA$ with $p\leq 2$. 
For example, the reaction $3A\stackrel{\tau}{\longrightarrow} \emptyset$ 
would lead to an additional contribution $-\tau(1-\hat{\phi}^3)\phi^3$ 
in the action~(\ref{UnshiftedUnrestricted1}), 
corresponding to the bare coupling constants
\begin{equation}
\Lambda_{1,3}=3\tau \,, \qquad 
\Lambda_{2,3}=3\tau \,, \qquad
\Lambda_{3,3}=\tau\,.
\end{equation}
Alternatively, the restriction may be incorporated directly in the 
field-theoretic action by inserting exponential damping factors, 
whose purpose is to `switch off' the reactions when a certain particle 
density is exceeded~\cite{Wijl01}. To this end each reaction term is 
multiplied by an exponential function $\exp(-nv\hat{\phi}\phi)$, 
where $n=\max(p,q)$ is the maximal number of particles involved 
in the respective reaction and $v$ is an additional coupling constant. 
When applied to the PCPD, one is led to the unshifted action
\begin{eqnarray}
\label{UnshiftedUnrestricted2}
S &= \int d^dx \int dt 
\biggl[&
\hat{\phi}(\partial_t-D\nabla^2)\phi - 
\mu (1-\hat{\phi}^2)\phi^2e^{-2v\hat{\phi}\phi} \\
&&- \nu (1-\hat{\phi})\hat{\phi}\phi^2e^{-2v\hat{\phi}\phi} + 
\sigma(1-\hat{\phi}) \hat{\phi}^2\phi^2e^{-3v\hat{\phi}\phi} \nonumber
\biggr].
\end{eqnarray}
Roughly speaking, the constant $v$ determines the field 
amplitude needed to declare location in space-time (representing a site) 
as occupied. Since the argument of the exponential function has to be
dimensionless and $[\hat{\phi}\phi]=\kappa^d$ the coupling constant 
$v$ is irrelevant, allowing the exponential functions to be expanded. 
To first order in $v$ this gives rise to the additional non-zero bare 
coupling constants
\begin{eqnarray*}
\label{BareCouplings2}
\Lambda_{1,3}&=&v(3\sigma-4\mu-2\nu) \\
\Lambda_{2,3}&=&v(9\sigma-6\mu-4\nu) \\
\Lambda_{3,3}&=&v(9\sigma-2\mu-2\nu) \\
\Lambda_{4,3}&=&3 \sigma v
\end{eqnarray*}
while higher-order contributions lead to additional non-vanishing bare 
coupling constants of the form $\Lambda_{p,q} \propto v^{q-2}$. 

In the inactive phase all terms of the leading series $\Lambda_{p,2}$ are 
marginal and thus have to be retained while all terms of higher order are 
irrelevant. In order to handle the series of marginal contributions, 
Wijland {\it et al} studied the corresponding generating function 
$G(x)=\sum_{p=1}^\infty \Lambda_{p,2}\phi^2\bar{\phi}^p$, recasting 
the flow equations of all coupling constants $\Lambda_{p,2}$ into a 
single functional renormalization group equation for the renormalized 
counterpart of $G$. Alternatively, one may work at the critical point, 
where the naive field dimensions of $\phi$ and $\bar{\phi}$ are equal, 
leading to a renormalizable field-theory~\cite{JanssenPrivate03}. 
Although a fixed point was shown to exist, it turned out to be 
unreachable from any physically meaningful initial condition. 
In addition, the fixed point itself is characterized by a negative 
variance of the order parameter fluctuations, which would have been 
unphysical as well. Finally, the critical exponents, in particular 
the exact result $z=2$, would have been incompatible with the 
existing numerical estimates which clearly indicate that $z<2$.

The failure suggest that a field-theoretic action in terms of a 
{\em single} coarse grained field $\phi$ (and its associated response 
field $\bar{\phi}$) may be incapable to describe the properties of the 
restricted PCPD. In fact, a single field for the density of particles 
may fail to describe the complex spatio-temporal interplay of particles 
and pairs, questioning the validity of the continuum limit on which 
Eq.~(\ref{UnshiftedUnrestricted2}) is based. Instead it may be more 
promising to devise a field-theory based on two different fields for 
single particles and pairs, separating the two dynamical modes in a 
similar way as in the cyclically coupled models of Sec.~\ref{CycSect}. 
However, a consistent formulation of such a field-theory is not yet known.

\subsection{On the impossibility of a Langevin equation for the PCPD}

Langevin equations are very popular in non-equilibrium statistical physics 
because of their intuitive simplicity and their enormous success in well-known 
systems such as DP. Therefore it has become customary to deal with 
phenomenological or even guessed Langevin equations even when a rigorous 
derivation is not available. For example, in the context of the PCPD several 
authors~\cite{Howa97,Park02a,Kock02,Hinr03a} have 
been discussing a Langevin equation of the form
\begin{equation}
\label{Langevin}
\!\!\!\!\!\!\!\frac{\partial}{\partial t}\rho(\xvec,t) = 
(\sigma-2\mu) \rho^2(\xvec,t) - c\rho^3(\xvec,t) + 
D\nabla^2\rho(\xvec,t) + \rho(\xvec,t)\xi(\xvec,t)\,,
\end{equation}
where $\xi(\xvec,t)$ is a white Gaussian noise with zero mean and 
the correlations
\begin{equation}
\label{Noise}
\langle \xi(\xvec,t)\xi(\xvec',t')\rangle = (4 \sigma - 2 \mu) \,
\delta^{(d)}(\xvec-\xvec') \delta(t-t') \,.
\end{equation}
This Langevin equation, if indeed realized, would be particularly appealing 
since it can be related to the problem of non-equilibrium 
wetting~\cite{Hinr03a}. However, as a mesoscopic description a Langevin 
equation assumes that all features of the process, in particular all 
particle-particle correlations, could be captured in terms of a 
single coarse-grained 
particle density. This works only in few cases (e.g. in DP), but not in the 
case of the PCPD, where anticorrelations produced by the 
annihilation/coagulation process play an important role.

In this context it is useful to recall that a Langevin equation can be derived 
from the field-theoretic action by transforming the terms of the form 
$\bar{\phi}^2 \phi^q$ into a noise and computing the functional derivative 
with respect to the response field $\bar{\phi}$. This procedure requires 
that terms involving cubic or higher-order powers of the response field are 
fully irrelevant. However, in the case of the PCPD cubic terms of the form 
$\bar{\phi}^3 \phi^q$ have to be retained in the action. 
Therefore, any attempt to describe the PCPD by a Langevin equation misses the 
physics of these terms and thus should lead to wrong results.
In addition, one would have to add an 
additional term $\xi_D(\xvec,t)$ on the r.h.s.
of the Langevin equation accounting for diffusive noise with the correlations
\begin{equation}
\langle \xi_D(\xvec,t)\xi_D(\xvec',t')\rangle = -D \, \rho(\xvec,t)
\nabla^2 \, \delta^{(d)}(\xvec-\xvec') \delta(t-t') \,
\end{equation}
which is expected to play an important role in the PCPD. 
However, it is at present not clear what
consequences the inclusion such an additional term would 
have~\cite{JanssenPrivate03}.

\subsection{Pair correlation functions in the unrestricted PCPD at criticality}

A different approach has been followed by Paessens and Sch\"utz \cite{Paes03b}
and leads to a non-trivial, unexpected result. They consider the 
{\em unrestricted} version of a generalized PCPD with the reactions
\begin{eqnarray}
mA \rightarrow (m+k)A  && \qquad \text{with rate }  \nu \nonumber \\
pA \rightarrow (p-l)A && \qquad  \text{with rate } \mu \\
A\emptyset \leftrightarrow\emptyset A && \qquad \text{with rate }\,  D 
\nonumber \ .
\end{eqnarray}
Models of this kind may be thought of as realizing a bosonic field-theory 
such as in Ref.~\cite{Howa97} and are specified by four integers $(m,p,k,l)$ 
and three rates $\nu,\mu$ and $D$. 
Paessens and Sch\"utz start from the master equation (\ref{mastereq}) and 
rewrite the quantum Hamiltonian $H$ in terms of space-time-dependent creation 
and annihilation operators $a^{\dag}(\vec{x},t)$ and $a(\vec{x},t)$ such that
the particle operator $n(\vec{x},t)=a^{\dag}(\vec{x},t)a(\vec{x},t)$. 
In what follows, the averages
\BEQ
\langle n(\vec{x},t)\rangle = \langle a(\vec{x},t)\rangle \;\; , \;\;
\langle n(\vec{x},t)^2\rangle = \langle a(\vec{x},t)^2\rangle 
+ \langle a(\vec{x},t)\rangle\;\; , \;\;
\EEQ
will be considered. 
Specializing to the case $m=p$, the critical line between the absorbing phase
and the active phase (with an infinite particle density) is located at
$\mu_c=\nu k/l$. For $\mu=\mu_c$, the spatially averaged 
particle density $|\Omega|^{-1}\int_{\Omega}\!\D\vec{x}\, 
\langle a(\vec{x},t)\rangle = \rho_0$ is a constant, where $\Omega$ is a 
spatial domain with volume $|\Omega|$. Remarkably, it turns out that at 
criticality and for $m=p$, the system of equations of motion for the variables 
$\{\langle a(\vec{x},t)\rangle,\langle a(\vec{x},t)a(\vec{y},t)\rangle\}$ 
closes if either $m=1$ or $m=2$ \cite{Paes03b}. 

The case $m=1$ is a `bosonic' version of the ordinary contact process and has
been considered as a model for biological clustering \cite{Houc02,Youn01}. 
It can be shown that the fluctuations $\langle a(\vec{x},t)^2\rangle$ of the
particle density diverge for $t\to\infty$ in dimensions $d\leq 2$, while they 
remain finite for $d>2$ \cite{Houc02,Paes03b}. 

The PCPD is described by the case $m=2$. After a time rescaling 
$t\mapsto t/(2D)$, and assuming translation-invariant initial conditions, 
consider the pair-correlator
\BEQ
F(\vec{r},t) := \langle a(\vec{x},t)a(\vec{x}+\vec{r},t)\rangle
\EEQ
(which is independent of $\vec{x}$) and the reduced coupling
\BEQ
\alpha := \frac{\nu}{D}\frac{k(k+l)}{2}
\EEQ
Then, under the stated conditions, the following equation holds on a 
hypercubic lattice
\BEQ \hspace{-10mm}
\frac{\partial}{\partial t}F(\vec{r},t) = \sum_{i=1}^{d} 
\left[ F(\vec{r}-\vec{e}_i,t) + F(\vec{r}+\vec{e}_i,t)
-2F(\vec{r},t)\right] +\delta_{\vec{r},\vec{0}}\,\alpha F(\vec{0},t)
\EEQ
where $\vec{e}_{i}$ is the unit vector in the $i^{\rm th}$ direction. 
If as initial condition one takes a Poisson distribution 
$F(\vec{r},0)=\rho_0^2$, one obtains the following Volterra integral
equation
\BEA
F(\vec{r},t) &=& \rho_0^2 + \alpha \int_{0}^{t} \!\D\tau\, b(\vec{r},t-\tau)
F(\vec{0},\tau) \\
b(\vec{r},t) &=& e^{-2dt} I_{r_1}(2t)\cdots I_{r_d}(2t)
\EEA
where $I_r$ is a modified Bessel function of order $r$. For $\vec{r}=\vec{0}$,
the same equation describes the kinetics of the mean spherical model, see
\cite{Godr00b,Pico02,Paes03a} but it is not yet clear if there is a deeper
reason for this relationship or whether it merely occurs by accident. 
The exact solution leads to the following long-time behaviour, at
$\mu=\mu_c$ \cite{Paes03b}
\BEQ \label{PS:Finfty}
F(\vec{0},t) \sim \left\{
\begin{array}{ll} \exp(t/\tau) & \mbox{\rm ~~;~  if $\alpha>\alpha_c$}\\
t^{\nu}                        & \mbox{\rm ~~;~  if $\alpha=\alpha_c$}\\
F_{\infty}                     & \mbox{\rm ~~;~  if $\alpha<\alpha_c$}
\end{array} \right.
\EEQ
where $\alpha_c$ is a known critical value such that $\alpha_c=0$ for $d\leq 2$
but it is finite for $d>2$, in addition 
$\tau=C_d\left(\frac{\alpha-\alpha_c}{\alpha}\right)^{-1/\nu}$, the
exponent $\nu=d/2-1$ for $2<d<4$ and $\nu=1$ for $d>4$, 
$F_{\infty}=\rho_0^2(1-\alpha/\alpha_c)^{-1}$ and $C_d$ is a known constant. 
As can be seen the two-point correlator diverges for large times if the
reduced rate $\alpha$ is large enough. 
This result can be generalized to yield $F(\vec{r},t)$ exactly, with a
similar conclusion. In particular, one obtains the dynamical exponent
$z=2$ \cite{Paes03b}. 

Since $\alpha$ measures the relative importance between the reaction and the
diffusion rates, the result (\ref{PS:Finfty}) 
means that for $d>2$ at high values of $D$ the 
distribution of the particle density along the critical line is relatively 
smooth and changes through a `clustering transition' (which occurs at a 
tricritical point) to a rough distribution for small values of $D$. 
This transition manifests itself in the variance 
$\sigma(t)^2 :=\langle n(\vec{x},t)^2\rangle
-\langle n(\vec{x},t)\rangle^2$ and not in the mean density
$\langle n(\vec{x},t)\rangle$ which is purely diffusive and has a 
constant average. 
A convenient order parameter of the clustering transition 
is $F_{\infty}^{-1}$ \cite{Paes03b}. 
On the other hand, for $d\leq 2$ there is a single transition along the 
critical line. 

These exact results for the bosonic PCPD \cite{Paes03b}
surprisingly agree with the conclusions of pair mean-field theory
and with the numerical results of \cite{Odor00} which suggested the
presence of two distinct 
transitions along the critical line. That scenario had, in the light of more 
recent simulations in $1D$, somewhat fallen into disfavour, see 
section~3. In fact, in
hindsight one observes that the existing cluster approximations of the
PCPD all explicitly restrict to clusters of a linear shape. At the level of
the pair approximation, at most two-site clusters need be considered and there
is no qualitative difference between one and several space dimensions. 
On the other hand, beginning with the triplet approximation, the geometrical
shape of the clusters also becomes important and one will have to distinguish
between linear clusters such as $\bullet\circ\bullet$ or  
$\bullet\bullet\bullet\circ$ (in $1D$ {\em only} such clusters occur) 
and non-linear ones as $\vekz{\bullet~~}{\circ\bullet}$ 
or $\vekz{\bullet\bullet~~}{~~\bullet\circ}$
which are meant to indicate more complicated geometrical forms as they will
only occur for $d\geq 2$. Indeed, truly three-dimensional clusters arise
for the first time in the quartet approximation ($N=4$). 
The triplet and higher approximations of the form as discussed in section~3 
only take {\em linear} clusters into 
account and should hence be expected to be more adapted to a truly
one-dimensional system than the pair approximation. Then a single transition
along the critical line should be expected for the $N\geq 3$ approximation 
while the different result of the
pair approximation might be related to the existence of two transitions for
$d>2$. 

It would be very interesting to see whether two transitions exist along the
critical line for the restricted PCPD in $d\geq 3$ dimensions. The interaction
terms in existing field-theory studies might be too close to the 
site-approximation to be able to see this and a study based on the master
equation is called for. 

\section{Possible generalizations}\label{ClassSec}

\subsection{Higher-order processes}

As shown in the preceding sections, the unusual critical behaviour of the PCPD 
is related to the fact that only {\em pairs} of particles can react while 
individual particle form a diffusing background. Therefore, as an obvious 
generalization, it is near at hand to study higher-order processes such as
\begin{equation}
nA \to (n+1)A \ , \quad nA \to \emptyset \ . \qquad  \qquad  (n\geq 3)
\end{equation}
Without diffusion, the number of  absorbing steady-states increases 
exponentially with the lattice size $L$. For example, for the triplet process
($n=3$) and in one dimension, we find
\BEQ \label{tcp:nst}
N_{\rm stat}(L) \simeq N_{0}\, g^L \;\; , \;\; 
N_0 = \left\{ \begin{array}{ll} 1           & \mbox{\rm ~~;~ periodic b.c.} \\
                                1.137\ldots & \mbox{\rm ~~;~ free b.c.}
\end{array} \right.
\EEQ
absorbing steady-states, where $g\simeq 1.839\ldots$ (see appendix~A). 
On the other hand, with single-particle diffusion,
only $n$ absorbing steady-states remain. 

The central question is whether such a system still exhibits a non-trivial 
phase transition. As a first na\"{\i}ve approach, the simple
mean-field equation
\begin{equation}
\label{GenMV}
\frac{\partial}{\partial t}\rho(\xvec,t) = b \rho^n(\xvec,t) - 
c\rho^{n+1}(\xvec,t) + 
D\nabla^2\rho(\xvec,t) 
\end{equation}
predicts the mean-field exponents
\begin{equation}
\label{MFExponents2}
\beta^{MF}=1,\quad 
\nuperp^{MF}=n/2,\quad
\nupar^{MF}=n.
\end{equation}
in sufficiently high space dimensions, including DP ($n=1$) and the 
PCPD ($n=2$) as special cases. The question posed here is whether the
higher-order processes, in particular the diffusive triplet-contact process 
(TCPD) with $n=3$, still displays a non-trivial critical behaviour in one 
spatial dimension.

The TCPD was introduced in \cite{Park02a}, where numerical 
simulations seemed to indicate a possible non-trivial behaviour at the 
transition. This point of view was confirmed by Kockelkoren and 
Chat{\'e}~\cite{Kock02}, who established the $1D$ TCPD as an 
independent universality class and with an upper critical dimension
$d_c=1$. However, {\'O}dor~\cite{Odor02d} questioned 
this claim by showing that the simulation results are compatible with 
mean-field exponents combined with appropriate logarithmic corrections 
(see table~\ref{TableTriplet}). 

Even more confusing are the findings for the quadruplet contact 
process $n=4$. In \cite{Kock02} the expected mean-field exponents are obtained 
while \'Odor finds numerical evidence 
for the presence of fluctuation effects~\cite{Odor02d}. The origin of these 
discrepancies is not yet understood. However, when comparing the results 
one should keep in mind that the model used by  Kockelkoren and Chat{\'e} 
is particularly suitable to simulate higher-order processes. 
In their model the reactions are local for any $n$, 
while in models with hard-core exclusion extended strings of $n$ 
particles have to line up before they can react. 

\begin{table}
\begin{center}
\begin{tabular}{|l|c|c|c|c|}
\hline Ref.                                    & $\delta$ & $z$      & $\beta$ & $\nupar$ \\ 
\hline Park~\etal~\cite{Park02a}               & 0.32(1)  & 1.75(10) & --      & 2.5(2) \\ 
\hline Kockelkoren and Chat{\'e}~\cite{Kock02} & 0.27(1)  & 1.8(1)   & 0.90(5) & --\\ 
\hline \'Odor~\cite{Odor02a}                   & 0.33(1)  & --       & 0.95(5) & --\\ 
\hline 
\end{tabular} 
\caption{
\label{TableTriplet}
Estimates for the critical exponents of the $1D$ triplet process TCPD.}
\end{center}
\end{table} 

\subsection{Towards a general classification scheme}

Extending these studies Kockelkoren and Chat{\'e}~\cite{Kock02} suggested that 
transitions in a  reaction-diffusion processes of the form
\begin{equation} \label{mnkl}
mA \to (m+k)A \ , \quad nA \to (n-l)A \ ,
\end{equation}
can be categorized according the the orders $m,n$ of creation and removal, 
while the numbers $k$ and $l$ determine additional symmetries such as parity 
conservation. In terms of $m,n$, they propose the general classification scheme 
shown in table~\ref{TableScheme}. In addition they investigate the r\^ole of 
parity-conservation in more detail, finding that such a symmetry does {\em not} 
alter the universality class whenever {\em every} sector includes an absorbing 
state. This conjecture is in agreement with the observation of 
Ref.~\cite{Park01} that an additional parity-conservation does not change the 
critical behaviour of the PCPD (see \cite{Odor03} for recent evidence of a PCPD
transition in the diffusive $2A\to 3A$, $4A\to\emptyset$ system, where
$m=2,n=4$). 

\begin{table}
\begin{center}
\begin{tabular}{|c|c|c|c|c|}
\hline  $m\backslash n$ & $\qquad 1\qquad$ & $\qquad 2\qquad$ & $\qquad 3\qquad$ & 
$\qquad 4\qquad$\\ 
\hline  1 & DP & DP/PC & DP   & DP   \\ 
\hline  2 & DP & PCPD  & PCPD & PCPD \\ 
\hline  3 & DP & DP    & TCPD & TCPD \\ 
\hline  4 & DP & DP    &  DP  & ?    \\ 
\hline 
\end{tabular} 
\end{center}
\caption{
\label{TableScheme}
General classification scheme of absorbing phase transitions in the $1D$ 
models (\ref{mnkl}) as proposed by Kockelkoren and 
Chat{\'e}~\cite{Kock02} and modified after \cite{ChatPrivat}.}
\end{table}

\section{Summary and Outlook}

In spite of intensive and prolonged efforts, the critical behaviour of
the one-dimensional PCPD is not yet understood. 
It is surprising to see that the
field-theoretical and numerical methods used to study the PCPD-transition have
led to such wildly different conclusions as listed in section~1. 
It appears that almost the only piece of information for which 
different methods obtain the
same result is the location $p_c(D)$ of the critical point (if applicable
at all). This continuing discrepancy is all the more astonishing because the
same techniques yield nicely consistent results when applied to models in the
DP or PC universality classes. By itself, this observation is a clear 
indication of a subtlety in the behaviour of the PCPD which is not present
in those other models. 

Specifically, at present there seems to be some majority opinion, based on
numerical work, in favour of a single new universality class along the
critical line and characterized by critical exponent values of the order
\BEQ
\delta \approx 0.2 \;\; , \;\; 
z \approx 1.7 - 1.8 < 2 \;\; , \;\;
\eta \approx 0.5 \;\; , \;\;
\beta/\nu_{\perp} \approx 0.5
\EEQ
On the other hand, a recent analytical study \cite{Paes03b} of the bosonic 
field-theory finds a dynamical exponent $z=2$ and a single 
universality class along the critical line, at least in one dimension. 
Their value of $z=2$ clearly is in disagreement with the numerical results 
{}from fermionic models. 
At the time of writing, it is not clear how this should be explained. 

Besides the majority opinion, several alternative scenarios have been
proposed. These include the possibility of more than one transition along
the critical line (which is known to be exact for the unrestricted bosonic
version in $d>2$ dimensions), continuously varying exponents and even the
cross-over to directed percolation after very long times. The last
possibility would mean that the diffusive background of solitary particles
would become irrelevant in the limit $t\to\infty$ which goes against the
current understanding of the bosonic field-theory. 

All in all, we are not (yet) able to decide between the several 
scenarios proposed. Pending further insight, the reader might find some comfort
in a quotation of `Thales'
\begin{quote}
Sei ruhig -- es war nur gedacht. \hfill {\em J.W. Goethe, Faust II (1830)}
\end{quote}
or alternatively with
\begin{quote}
It were not best that we should all think alike; it is the difference of
opinion which makes horseraces. \hfill {\em M. Twain, Pudd'nhead Wilson (1894)}
\end{quote}
We look forward to future insightful studies shedding more light on the so 
simple-looking PCPD. 

\vspace{10mm} \noindent
{\bf Acknowledgments}

\noindent
It is a pleasure to thank G. Barkema, E. Carlon, H. Chat\'e, P. Grassberger, 
M. Howard, H.K. Janssen, J.F.F. Mendes, M.A. Mu\~noz, G. \'Odor, H. Park,
U. Schollw\"ock, G.M. Sch\"utz, U. T\"auber, F. van Wijland 
for their collaboration/discussions/correspondence in trying to understand the 
behaviour of the PCPD. We thank S. L{\"u}beck and M. Pleimling for 
critical readings of the manuscript, R. Dickman and G. \'Odor for help with 
some figures and the Max-Planck Institut f\"ur Physik komplexer Systeme (MPIPKS)
in Dresden for warm hospitality during the NESPHY03 Seminar, where part of this 
work was done. 


\appsection{A}{The absorbing states in the pair-contact process}

Consider the $1D$ pair-contact process on a lattice with $L$ sites. We derive
the number of absorbing steady-states $N_{\rm stat}(L)$ given in 
eq.~(\ref{pcp:nst}), following \cite{Carl01}. The main ingredient is that any 
state which only contains isolated particles is an absorbing steady-state. 

First, we consider free boundary conditions. Obviously, $N_{\rm stat}(1)=2$
and $N_{\rm stat}(2)=3$. Now fix the leftmost site. If it is occupied, its
neighbour must be empty in order to obtain a steady-state and it remains to
consider an open chain of $L-2$ sites. On the other hand, if the leftmost site
is empty, one considers the remaining open chain of $L-1$ sites. We thus
have the recursion
\BEQ
N_{\rm stat}(L) = N_{\rm stat}(L-2) + N_{\rm stat}(L-1)
\EEQ
Because of the initial conditions,
\BEQ
N_{\rm stat}(L) = F_{L+1} = \frac{g_+^{L+2} - g_-^{L+2}}{g_+ - g_-}
\EEQ
is the $(L+1)^{\rm th}$ Fibonacci number, where $g_{\pm}=(1\pm\sqrt{5})/2$. 

On the other hand, for periodic boundary conditions, we fix one of the sites.
If that site is occupied, both its left and right nearest neighbours must be
empty and we are left with an {\it open} chain of $L-3$ sites. But if that site
is empty, we are left with an open chain of $L-1$ sites. Therefore
\BEQ
N_{\rm stat}^{({\rm per})}(L) = N_{\rm stat}(L-3) + N_{\rm stat}(L-1) 
= g_+^L +  g_-^L
\EEQ
{}From these results we recover (\ref{pcp:nst}) for $L$ sufficiently large. 

In general, $N_{\rm stat}(L)$ is found from its generating function
${\cal G}(z) = \sum_{L=0}^{\infty} N_{\rm stat}(L) z^L$. For the kind 
of application at hand, ${\cal G}(z)=z(2+z)/(1-z-z^2)$ is a simple rational 
function of $z$ and its coefficients can be obtained in a closed form as 
follows. 

\noindent {\bf Lemma:} {\it Let $P(z)$ and $Q(z)$ be polynomials of orders
$p$ and $q$, respectively and let $p<q$. If one has in addition
$Q(z)=q_0 (1-z\rho_1)\cdots (1-z\rho_q)$ such that the $\rho_i$ are pairwise
distint for $i=1,\ldots,q$ and $q_0$ is a constant, then}
\BEA
f(z) &=& \frac{P(z)}{Q(z)} = \sum_{n=0}^{\infty} f_n z^n 
\nonumber \\
f_n  &=& - \sum_{j=1}^{q} \frac{\rho_j^{n+1} P(1/\rho_j)}{Q'(1/\rho_j)}
\EEA

In \cite{Grah94} this is proven in an elementary way, 
here we give a function-theory proof which easily generalizes to the case 
when $Q(z)$ has multiple zeroes. We have
\begin{displaymath}
f_n = \frac{1}{n!} f^{(n)}(0) = \frac{1}{2\pi\II} \oint_{\cal C}
\D w\, \frac{f(w)}{w^{n+1}} 
= \frac{1}{2\pi\II}\oint_{{\cal C}'}\D u\, \frac{u^n P(1/u)}{u\, Q(1/u)}
\end{displaymath}
where we have set $u=1/w$. The contour $\cal C$ 
is a circle around the origin with
a radius smaller than the convergence radius of $f(z)$ and ${\cal C}'$ encloses
all the points $\rho_j$, $j=1,\ldots,q$. Since $p<q$, there is no singularity
at $u=0$ and the only singularities of $u^{-1}f(1/u)$ are the simple p\^oles 
located at $u=\rho_j$. If we concentrate on the singularity at $u=\rho_1$, we 
have $u Q(1/u)=(u-\rho_1) q_0 \prod_{j=2}^{q} \left(1-\rho_j/u\right)$ such 
that the product is regular at $u=\rho_1$. On the other hand, the derivative
$Q'(1/\rho_1)= -\rho_1 q_0 \prod_{j=2}^{q} \left(1-\rho_j/\rho_1\right)$.
Therefore, close to $u\simeq\rho_1$, we have 
\begin{displaymath} 
uQ(1/u)\simeq -(u-\rho_1) \frac{Q'(1/\rho_1)}{\rho_1}
\end{displaymath}
{}From the residue theorem and summing over all simple poles of $f(z)$ the
assertion follows. \hfill q.e.d. 

Other processes can be treated similarly. As an example, consider the
triplet-contact process without diffusion as introduced in section~7.1. 
Absorbing states are those without any triplet of neighbouring occupied sites. 
On a $1D$ lattice with free boundary conditions and $L$ sites, we have the
recursion 
\BEQ
N_{\rm stat}(L) = N_{\rm stat}(L-1)+ N_{\rm stat}(L-2) + N_{\rm stat}(L-3)
\EEQ
together with the initial conditions $N_{\rm stat}(1)=2$, $N_{\rm stat}(2)=4$ 
and $N_{\rm stat}(3)=7$. Therefore the generating function is 
${\cal G}(z)=z(2+2z+z^2)/(1-z-z^2-z^3)$ and its coefficients are easily
read off. For periodic boundary conditions, we find 
\BEQ
N_{\rm stat}^{({\rm per})}(L) = N_{\rm stat}(L-1) + N_{\rm stat}(L-3) 
+ 2N_{\rm stat}(L-4) 
\EEQ
For $L$ sufficiently large, we thus arrive at eq.~(\ref{tcp:nst}). 

\appsection{B}{On finite-size scaling techniques}

In systems like the PCPD, where the the critical line separates a non-critical
ordered phase from a critical disordered phase, the analysis of finite-size
data needs some modifications with respect to the usual situation, where both
the ordered and the disordered phases are non-critical. To be specific, we 
shall discuss here the finite-size scaling of the lowest gap $\Gamma$ of the
quantum Hamiltonian $H$, see section~\ref{sect4}, and following \cite{Carl01}.
As usual, we begin with the finite-size scaling form
\BEQ
\Gamma_L(p) = L^{-z} f\left( (p-p_c)L^{1/\nu_{\perp}}\right)
\EEQ
where $p$ is the control variable which measures the distance to the
critical point $p_c$ (the dependence on $D$ is suppressed throughout) and
$f$ is assumed to be continously differentiable. One
expects the asymptotic behaviour, see eq.~(\ref{GammaL})
\BEQ \label{B:gl:GammaL}
\Gamma_L(p) \sim \left\{\begin{array}{ll} 
e^{\sigma L} & \mbox{\rm ~~;~ if $p<p_c$} \\
L^{-2}       & \mbox{\rm ~~;~ if $p>p_c$}
\end{array}\right. 
\EEQ
where $\sigma=\sigma(p)$ is a constant. In the usual case, one would have
instead $\Gamma_L(p)\sim \Gamma_{\infty}$ for $p>p_c$. From (\ref{B:gl:GammaL}),
one finds for the scaling function
\BEQ
f(\mathfrak{z}) \sim \left\{\begin{array}{ll}
\exp(-A |\mathfrak{z}|^{\nu_{\perp}}) 
& \mbox{\rm ~~;~ if $\mathfrak{z}\to -\infty$} \\
\mathfrak{z}^{ (z-2)\nu_{\perp}}
& \mbox{\rm ~~;~ if $\mathfrak{z}\to +\infty$} 
\end{array}\right. 
\EEQ
where $A$ is a positive constant. Therefore, since $f(\mathfrak{z})>0$, it 
follows that for $z<2$ the scaling function $f$ must have a maximum at some
finite value $\mathfrak{z}_{\rm max}$. Next, the logarithmic derivative
(\ref{logderGamma}) becomes
\BEQ
Y_L = -z+\frac{\ln[ f(\mathfrak{z}_+)/f(\mathfrak{z}_-)]}{\ln[(L+1)/(L-1)]}
\EEQ
where $\mathfrak{z}_{\pm}=(p-p_c)(L\pm 1)^{1/\nu_{\perp}}$. Furthermore, 
it is easy to see that
in the scaling limit $p\to p_c$ and $L\to\infty$ such that 
$\mathfrak{z}=(p-p_c)L^{1/\nu_{\perp}}$ is kept fixed, one has
\BEQ
\lim \frac{\D Y_L}{\D p} \simeq \left\{\begin{array}{ll}
L^{1/\nu_{\perp}} A(2-\nu_{\perp}) (-\mathfrak{z})^{\nu_{\perp}-1}
& \mbox{\rm ~~;~ if $\mathfrak{z}\to -\infty$} \\
L^{1/\nu_{\perp}} (z-2) \mathfrak{z}^{-1}
& \mbox{\rm ~~;~ if $\mathfrak{z}\to +\infty$} 
\end{array}\right. 
\EEQ
Provided that $z<2$ and $\nu_{\perp}<2$, 
there {\em must} exist a finite $\mathfrak{z}^*$ such that
$\left.\D Y_L/\D p\right|_{\mathfrak{z}=\mathfrak{z}^*}=0$. However, since
\BEQ
Y_L(\mathfrak{z}^*) = -z +\frac{1}{\nu_{\perp}} 
\frac{\mathfrak{z}^* f'(\mathfrak{z}^*)}{f(\mathfrak{z}^*)}
\EEQ
that maximum value of $Y_L$ {\em cannot} be used to estimate the dynamical
exponent $z$ \cite{Carl01}. Rather, one has to form first a sequence of
estimates $p_L$ of the critical point $p_c$ from the above extremum criterion
which should converge according to 
$p_L \simeq p_c + \mathfrak{z}^* L^{-1/\nu_{\perp}}$. Having found $p_c$,
estimates of $z$ can finally be obtained from eq.~(\ref{4:gl:Exponenten}). 

%
%
\begin{figure}
\centerline{\includegraphics[width=120mm]{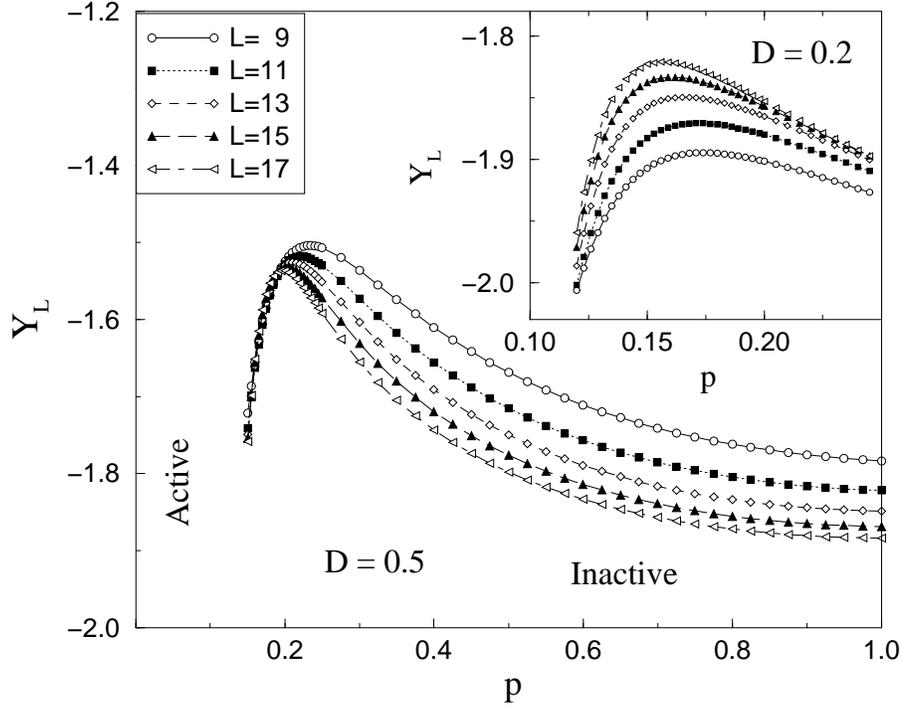}}
\caption{\label{abb:YL} Plot of the function $Y_L(p)$ in the $1D$ PCPD 
as a function of $p$ for several lattice size $L$ and $D=0.5$. The inset shows 
the case $D=0.2$. After {\protect\cite{Carl01}}.
}
\end{figure}
%
%

We finally point out that the habitual method of intersecting two curves
$Y_L(p)$ and $Y_{L'}(p)$ for two lattice sizes $L,L'$ does not work in 
general. For example, it is known for the $1D$ PCPD with free boundary 
conditions, that the curves $Y_L(p)$ do not intersect for different values
of $L$ \cite{Carl01}. We illustrate this in figure~\ref{abb:YL}.

\appsection{C}{On universal amplitudes in nonequilibrium criticality}

We recall the phenomenological scaling arguments about universal
scaling amplitudes in non-equilibrium critical points and derive
eqs.~(\ref{Ampxi},\ref{Ampr}) for sytems in the directed percolation
universality class, following \cite{Henk01b}. We shall denote time by 
$r_{\|}$ and space by $r_{\perp}$. The distance from the steady-state
critical point is mesured by $t$ and $h$ denotes an external field 
(e.g., for directed percolation $t=p-p_c$ and $h$ is the rate of a 
process $\emptyset\to A$). We shall assume translation invariance
throughout. The arguments involves tracing the non-universal metric
factors, generalizing a similar by now classic line of thought for
equilibirium systems \cite{Priv84}. 
This is most conveniently done in the bulk first before finally introducing
lattices of finite extent. 

Physical quantities of interest are the mean particle 
density $\rho$, the survival probability $P$ and the pair connectedness 
function $G=G(r_{\perp}',r_{\|}';r_{\perp},r_{\|})$, which is defined as the
probability that the sites $(r_{\perp}',r_{\|}')$ and $(r_{\perp},r_{\|})$ are
connected by a direct path \cite{Marr99,Hinr00}. 
Because of translation invariance $G=G(r_{\perp}'-r_{\perp},r_{\|}'-r_{\|})$. 
These quantities are expected to satisfy the scaling behaviour
\BEA
\hspace{-20truemm} \rho(r_{\perp},r_{\|};t,h) &=& 
b^{-x_{\rho}} \rho\left( 
\frac{r_{\perp}}{b},\frac{r_{\|}}{b^z}; t b^{y_t}, h b^{y_h}\right) 
=D_{1\rho}\,\xi_{\perp}^{-x_{\rho}} 
{\cal E}^{\pm}\left( \frac{r_{\perp}}{\xi_{\perp}},
D_0 \frac{r_{\|}}{\xi_{\perp}^z}; D_2 h |t|^{-y_h/y_t}\right) \nonumber\\
\hspace{-20truemm} P(r_{\perp},r_{\|};t,h) &=& 
b^{-x_{P}} P\left( 
\frac{r_{\perp}}{b},\frac{r_{\|}}{b^z}; t b^{y_t}, h b^{y_h}\right) 
=D_{1P}\, \xi_{\perp}^{-x_{P}} 
{\cal F}^{\pm}\left( \frac{r_{\perp}}{\xi_{\perp}},
D_0 \frac{r_{\|}}{\xi_{\perp}^z}; D_2 h |t|^{-y_h/y_t}\right) \label{gl:rPG}
\nonumber \\
\hspace{-20truemm} G(r_{\perp},r_{\|};t,h) &=& 
b^{-x_{G}} G\left( 
\frac{r_{\perp}}{b},\frac{r_{\|}}{b^z}; t b^{y_t}, h b^{y_h}\right) 
=D_{1G}\, \xi_{\perp}^{-x_{G}} {\cal G}^{\pm}
\left( \frac{r_{\perp}}{\xi_{\perp}},
D_0 \frac{r_{\|}}{\xi_{\perp}^z}; D_2 h |t|^{-y_h/y_t}\right) \nonumber \\
 & &  \label{C1}
\EEA
where the $x$'s are scaling dimensions and $y_{t,h}$ renormalization group
eigenvalues (in particular $y_t=1/\nu_{\perp}$), the $D$'s are 
non-universal metric factors, ${\cal E},{\cal F},{\cal G}$ are universal
scaling functions where the index distinguishes between the cases $t>0$ and
$t<0$, $\xi_{\perp}=\xi_0 |t|^{-\nu_{\perp}}$ is the spatial, 
$\xi_{\|}=\xi_{\perp}^z /D_0$ is the temporal correlation length and $z$ is
the dynamical exponent. 

In the steady state, and for $h=0$, one expects $\rho\sim t^{\beta}$ and
$P\sim t^{\beta'}$. In general, the two exponents $\beta$ and $\beta'$ are 
distinct from each other. For spatial translation invariance, the dependence
on $r_{\perp}$ drops out for both $\rho$ and $P$ and in the steady state
(i.e. $r_{\|}\to\infty$) one has
\BEA
\rho(t,h) &=& D_{1\rho}\,\xi_{0}^{-\beta/\nu_{\perp}} 
\wit{\cal E}^{\pm}\left(  D_2 h |t|^{-y_h/y_t}\right) |t|^{\beta} \nonumber\\
P(t,h) &=& D_{1P}\,\xi_{0}^{-\beta'/\nu_{\perp}} 
\wit{\cal F}^{\pm}\left(  D_2 h |t|^{-y_h/y_t}\right) |t|^{\beta'}
\label{gl:DichteWkeit} 
\EEA
where $x_{\rho}=\beta/\nu_{\perp}$, $x_{P}=\beta'/\nu_{\perp}$ 
and $\wit{\cal E}^{\pm}=\lim_{r_{\|}\to\infty} {\cal E}^{\pm}$ 
and similarly for $\cal F$. We also consider the auto-connectedness (that is 
$r_{\perp}=r_{\perp}'$) in the steady state
\BEQ 
G(0,\infty;t,h) =: G(t,h) = D_{1P}\,\xi_{0}^{-x_G} 
\wit{\cal G}^{\pm}\left(  D_2 h |t|^{-y_h/y_t}\right) |t|^{x_G \nu_{\perp}}
\EEQ
In the active phase ($t>0$), the surviving clusters will create an average
density $\sim |t|^{\beta}$ in the interior of the spreading cone. Therefore,
the auto-connectedness should in the steady state saturate at the value
\cite{Mend94}
\BEQ \label{gl:GrP}
G(t,h) = \rho(t,h) P(t,h)
\EEQ 
Comparison of the scaling forms then yields, setting $h=0$,
\BEQ \label{gl:GrhoP}
x_G = (\beta+\beta')/\nu_{\perp} \;\; , \;\;
D_{1G} = D_{1\rho}D_{1P}\, 
\frac{\wit{\cal E}^{\pm}(0) \wit{\cal F}^{\pm}(0)}{\wit{\cal G}^{\pm}(0)}
\EEQ
Usually, $x_G=d-\theta z$ is expressed in terms of the initial critical
slip exponent $\theta$ \cite{Jans89}, 
which makes it apparent that the expression (\ref{gl:GrhoP}) is
in fact a generalized hyperscaling relation, see \cite{Mend94}. 

Next, we consider the total mass $M$ of the cluster, given by
\BEQ\label{gl:Masse}
\hspace{-8truemm} 
M(t,h) := \int_{\rm I\!R^d} \!\D^d r_{\perp} \int_{0}^{\infty} \!\D r_{\|}\,
G(r_{\perp},r_{\|};t,h) = \frac{D_{1G}}{D_0} \xi_{\perp}^{\gamma/\nu_{\perp}} 
\overline{\cal G}^{\pm}\left( D_2 h |t|^{-y_h/y_t}\right)
\EEQ
where eq.~(\ref{gl:rPG}) was used and $\overline{\cal G}^{\pm}$ is a new 
universal function related to ${\cal G}^{\pm}$. Also
\BEQ \label{gl:hypgamma}
\gamma = d\nu_{\perp} + \nu_{\|} - \beta -\beta' 
\EEQ
which is the analogue of the hyperscaling relation of equilibrium systems. 

While the discussion so far has been completely general, we now appeal to
two properties which are valid for systems in the directed percolation
universality class, but need not be generically valid. First, we consider
a directed percolation process in the presence of a weak field $h$
(physically, $h$ parametrises the rate of a particle creation process
$\emptyset\to A$). A site
at a given time becomes active if it is connected with at least one active site
in the past, where a particle was created by the field. The number of such
sites is equal to the cluster size, the probability to become active
is given by the density 
\BEQ
\rho(t,h) \simeq 1 - (1-h)^{M(t,h)} \simeq h M(t,h)
\EEQ
for $h$ small. Therefore, 
\BEQ \label{gl:Msuszep}
M(t,0) = \left. \frac{\partial \rho(t,h)}{\partial h}\right|_{h=0}
\EEQ
Comparison with the scaling forms for $\rho$ and $M$ leads to
\BEQ \label{gl:DDD}
y_h/y_t = \beta+\gamma \;\; , \;\;
D_{1P} = D_0 D_2 \xi_{0}^{-(\beta+\gamma)/\nu_{\perp}} {\cal A}^{\pm}
\EEQ
where ${\cal A}^{\pm}$ is an universal amplitude. Second, directed percolation
is special in the sense that there is a `duality' symmetry which 
can be used to show that \cite{Gras79}
\BEQ \label{gl:dual}
\rho(t,h) = P(t,h)
\EEQ
As a consequence, $\beta=\beta'$ and $D_{1\rho}=D_{1P}$ for directed 
percolation and we thus have, combining 
eqs.~(\ref{gl:DichteWkeit},\ref{gl:Masse},\ref{gl:DDD})
\BEA
\rho(t,h) &=& D_0 D_2 \xi_{0}^{-d-z} |t|^{\beta} 
\hat{\cal M}_1^{\pm}\left( D_2 h |t|^{-\beta-\gamma}\right)
\nonumber \\
M(t,h) &=& D_0 D_2^2 \xi_{0}^{-d-z} |t|^{-\gamma} 
\hat{\cal M}_2^{\pm}\left( D_2 h |t|^{-\beta-\gamma}\right)
\EEA
with universal functions $\hat{\cal M}_n^{\pm}(x)
=\D^n \hat{\cal M}^{\pm}(x)/\D x^n$ and where the hyperscaling relation 
eq.~(\ref{gl:hypgamma}) has been used.    
Finally, we define a new function $\mu=\mu(t,h)$ by 
$\rho(t,h) = \partial \mu(t,h)/\partial h$, which implies
\BEQ
\mu(t,h) = D_0 \xi_{0}^{-d-z} |t|^{(d+z)\nu_{\perp}} \hat{\cal M}^{\pm}
\left( D_2 h |t|^{-\beta-\gamma}\right)
\EEQ
In particular we have because of
$\xi_{\|}=\xi_{\perp}^z/D_0$ that
\BEQ \label{gl:univmu}
\mu(t,0) \xi_{\perp}^d(t,0) \xi_{\|}(t,0) \mathop{\rar}_{t\to0} 
\mbox{\rm univ. constant},
\EEQ
again quite analogous to an equilibrium result, see \cite{Priv84}.

At last, we consider a geometry of finite size $L$ in space but of infinite
extent in time. As usual in finite-size scaling \cite{Priv84}, 
we postulate that in this finite geometry merely the scaling functions 
are modified
\BEQ
\hat{\cal M}_n^{\pm} = \hat{\cal M}_n^{\pm} 
\left( D_2 h |t|^{-\beta-\gamma}; L\xi_{\perp}^{-1}\right) 
\EEQ
and without introducing any further metric factor. 
Indeed, we can then scale out $L$ and, because of eq.~(\ref{gl:univmu}),
arrive at the scaling forms (\ref{Ampxi}). The universality
of the moment ratio (\ref{Ampr}) is obtained from the analogous extension of 
(\ref{C1}) as follows 
\BEA
\langle r_{\perp}^n\rangle
&=& \frac{\int_{\Omega(L)}\!\D^d r_{\perp}\int_{0}^{\infty}\!\D r_{\|}\, 
r_{\perp}^n G(r_{\perp},r_{\|};L/\xi_{\perp})}
{\int_{\Omega(L)}\!\D^d r_{\perp}\int_{0}^{\infty}\!\D r_{\|}\, 
G(r_{\perp},r_{\|};L/\xi_{\perp})} \nonumber\\
&=& \xi_{\perp}^n
\frac{\int_{\Omega(L/\xi_{\perp})}
\!\D^d r_{\perp}\int_{0}^{\infty}\!\D r_{\|}\,
r_{\perp}^{n-x_G} {\cal G}^{\pm}(r_{\perp},r_{\|};L/\xi_{\perp})}
{\int_{\Omega(L/\xi_{\perp})}\!\D^d r_{\perp}\int_{0}^{\infty}\!\D r_{\|}\,
r_{\perp}^{-x_G} {\cal G}^{\pm}(r_{\perp},r_{\|};L/\xi_{\perp})}
\nonumber \\
&=& \xi_{\perp}^n \bar{\Xi}_{n}(L/\xi_{\perp})\nonumber \\
&=& L^n {\Xi}_{n}(L/\xi_{\perp})
\EEA
where $\Omega(L)$ is a $d$-dimensional hypercube of linear extent $L$ and 
$\bar{\Xi}_n$ and ${\Xi}_n$ are universal functions. 
   
\newpage 


\vspace{10mm} \noindent
{\bf \large References}\\[5mm]
\bibliographystyle{jpa}		
\footnotesize			
\baselineskip 0mm		
\bibliography{revue_PCPD}       
\end{document}